\documentclass[amsmath,amssymb,twocolumn,superscriptaddress, prl]{revtex4-2}
\usepackage{physics}
\usepackage{bm}
\usepackage{graphicx}
\usepackage{xcolor}
\usepackage{verbatim}
\usepackage{MnSymbol}
\usepackage[colorlinks=true, breaklinks=true, linkcolor=red, citecolor=blue, urlcolor=blue]{hyperref}

\begin{document}
\title{Twisted chiral superconductivity in photodoped frustrated Mott insulators}
\author{Jiajun Li}
\affiliation{Laboratory for Theoretical and Computational Physics, Paul Scherrer Institute, 5232 PSI Villigen, Switzerland}
\affiliation{Department of Physics, University of Fribourg, 1700 Fribourg, Switzerland}
\author{Markus M\"{u}ller}
\affiliation{Laboratory for Theoretical and Computational Physics, Paul Scherrer Institute, 5232 PSI Villigen, Switzerland}
\author{Aaram J. Kim}
\affiliation{Department of Physics, University of Fribourg, 1700 Fribourg, Switzerland}
\author{Andreas M. L\"{a}uchli}
\affiliation{Laboratory for Theoretical and Computational Physics, Paul Scherrer Institute, 5232 PSI Villigen, Switzerland}
\affiliation{Institute of Physics, \'{E}cole Polytechnique F\'{e}d\'{e}rale de Lausanne (EPFL), 1015 Lausanne, Switzerland}
\author{Philipp Werner}
\affiliation{Department of Physics, University of Fribourg, 1700 Fribourg, Switzerland}

\begin{abstract}

Recent advances in ultrafast pump-probe spectroscopy provide access to hidden phases of correlated matter, including light-induced superconducting states, but the theoretical understanding of these nonequilibrium phases remains limited. Here we report how a new type of chiral superconducting phase can be stabilized in photodoped frustrated Mott insulators. 
The metastable phase features a spatially varying order parameter with a $120^\circ$ phase twist which breaks both time-reversal and inversion symmetry. Under an external electric pulse, the $120^\circ$ chiral superconducting state can exhibit a second-order supercurrent perpendicular to the field in addition to a first-order parallel response, similar to a nonlinear anomalous Hall effect. This phase can be tuned by artificial gauge fields when the system is dressed by high-frequency periodic driving. The mechanism revealed in this study applies to Mott insulators on various frustrated lattices and the hidden superconducting phase can be realized in both cold-atom quantum simulators and correlated solids. 

\end{abstract}
\maketitle

The last decades have witnessed an intensive search for superconducting states with spontaneously broken time-reversal and inversion symmetry. Historically the first proposal involved chemically doped Mott insulating states of high-$T_c$ copper oxides, although the predicted properties have not been detected in experiments 
\cite{kiefl1990}. Since then the tantalizing chiral superconducting state, an essential building block for topological quantum computers, 
has been sought after in various systems, such as Sr$_2$RuO$_4$ \cite{mackenzie2003}, UPt$_3$ \cite{joynt2002,avers2020} and lately also in twisted bilayer graphene \cite{cao2016}. In the meantime, advances in cold-atom experiments and ultrafast spectroscopy have enabled the exploration of metastable hidden states of correlated systems, which are not accessible via a thermal pathway \cite{ichikawa2011,stojchevska2014,depaz2013,singh2019}. A 
remarkable phenomenon 
is the putative light-induced superconductivity, which has been observed in the vicinity of Mott insulating phases \cite{fausti2011, mitrano2016, buzzi2020, budden2021}. One promising route to such hidden phases is photodoping, where the laser excitation drastically redistributes charges by simultaneously creating long-lived particle-like (doublon) and hole-like (holon) charge carriers \cite{iwai2003,okamoto2010,beaud2014,dean2016,ligges2018}. With low enough entropy, the photocarriers may form a BEC-like superconducting condensate \cite{buzzi2020} supported by the doublon-holon exchange mediated by virtual charge recombination processes \cite{rosch2008,peronaci2020,li2020,murakami2022}. It is   
thus tempting 
to ask whether a chiral superconducting state can be realized as a hidden phase in photodoped Mott insulators.

Here we use state-of-the art numerical simulations to demonstrate that a chiral superconducting state with a $120^\circ$ phase twist can be stabilized by photodoping Mott insulators on frustrated lattices. The 
non-equilibrium nature of photodoped quasiparticles endows them with 
an exchange term of positive sign, 
an important ingredient favoring chiral condensates especially on frustrated lattices \cite{kalmeyer1987}. The $120^\circ$ order has been discussed in noncollinear antiferromagnets and Floquet-driven bosonic systems \cite{struck2013}. Here instead we consider a fermionic condensate exhibiting an exotic supercurrent response that requires no or little Floquet engineering. 
Remarkably, 
the $120^\circ$ superconducting phase features a spatially varying order parameter and persistent currents which change handedness under the inversion or time-reversal transformation, even in the absence of an external magnetic field. The condensate is found to prevail in an extended parameter regime, and can be further enhanced by artificial gauge fields created by periodic driving \cite{struck2012,hauke2012,claassen2017}. We confirm the existence of this ordered hidden phase on a Bethe lattice with infinite coordination number using nonequilibrium dynamical mean-field theory, and on the triangular and Kagome lattices using exact diagonalization. 
The 
exotic order 
may be realized 
in fermionic cold-atom systems and Kagome-lattice correlated materials, such as herbertsmithite \cite{shores2005}, where its stability can be tuned through Floquet-engineered artificial gauge fields. It may also be relevant for the light-induced superconducting state observed in $\kappa$--(BEDT--TTF)$_2$Cu[N(CN)$_2$]Br \cite{buzzi2020}.

\begin{figure*}[t]
\includegraphics[scale=0.38]{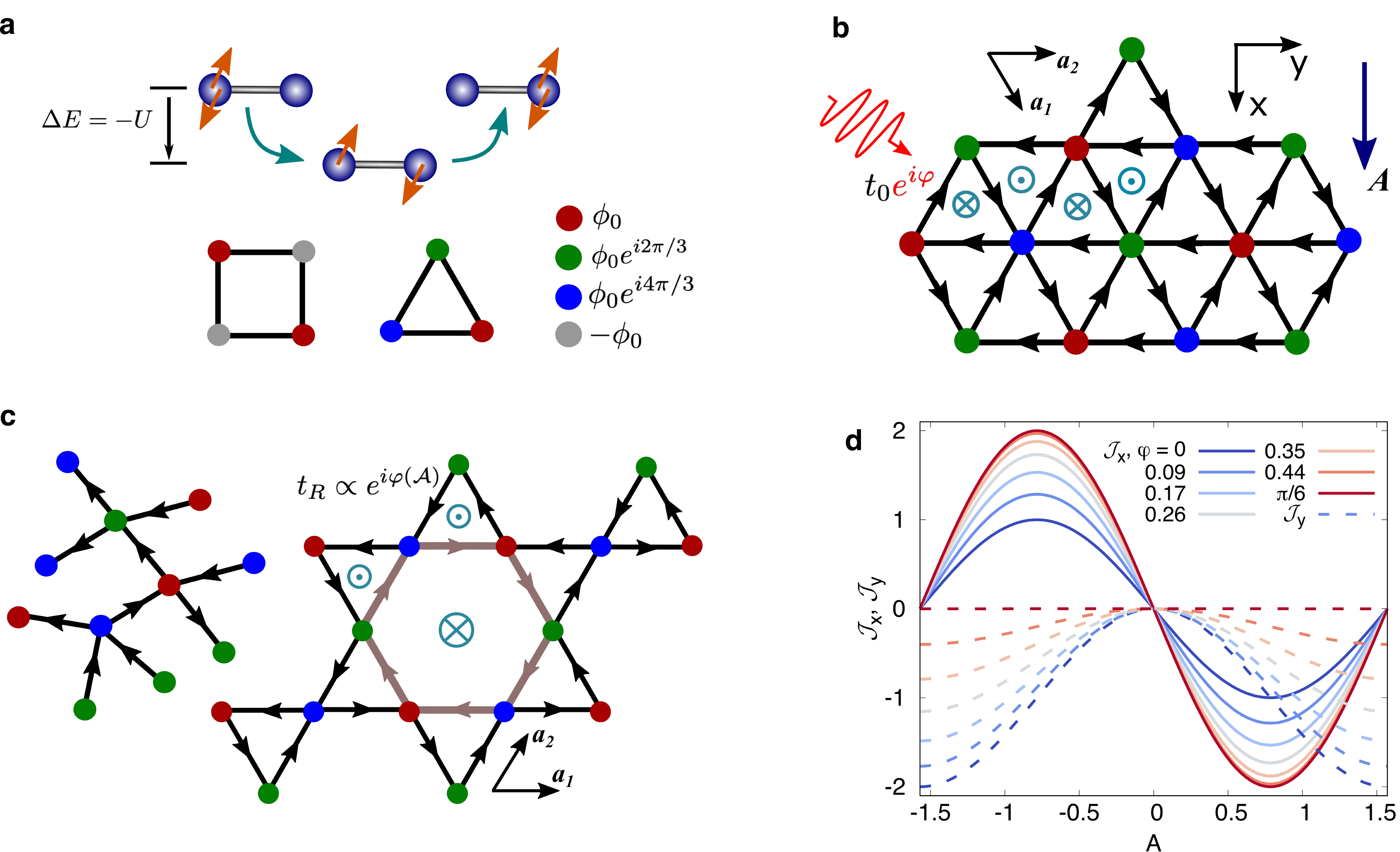}
\caption{{\bf Doublon-holon condensate on a three-colorable frustrated lattice.} The persistent currents are indicated by black arrows. A high-frequency driving (red wavy line in panel (b)) induces an artificial gauge field coupling to the hopping $t_0$ through the Peierls phase $\varphi$, resulting in a flux (indicated by a dot and cross) for the triangular and hexagonal motifs in panels (b) and (c). (a) The doublon-holon exchange interaction mediated by a virtual recombination. The amplitude $J_{\perp}$ is positive, in contrast to analogous processes in the equilibrium attractive Hubbard model. It can induce staggered $\eta$--pairing on bipartite lattices and $120^\circ$ pairing on three-colorable lattices. (b) The condensate on the triangular lattice. $\bm a_{1,2}$ are the basis vectors of the lattice. This panel also shows a constant vector potential along the $x$--direction. (c) The condensates on the Bethe (left) and Kagome (right) lattices. One of several possible $120^\circ$ condensates (the $\bm q =0$ order) is shown for the Kagome lattice. The magnitude of the flux through the hexagon is twice that of a triangle. (d) The longitudinal ($\mathcal{J}_x$, solid lines) and transverse ($\mathcal{J}_y$, dashed lines) superconducting current density under the constant vector potential $\bm A$ along the $x$--direction for the triangular lattice (with $A=2|\bm A|\cos(\pi/6)$), see panel (b). $\varphi$ is given in radians.
}
\label{diagram}
\end{figure*}

\begin{figure*}[t]
\includegraphics[scale=0.7]{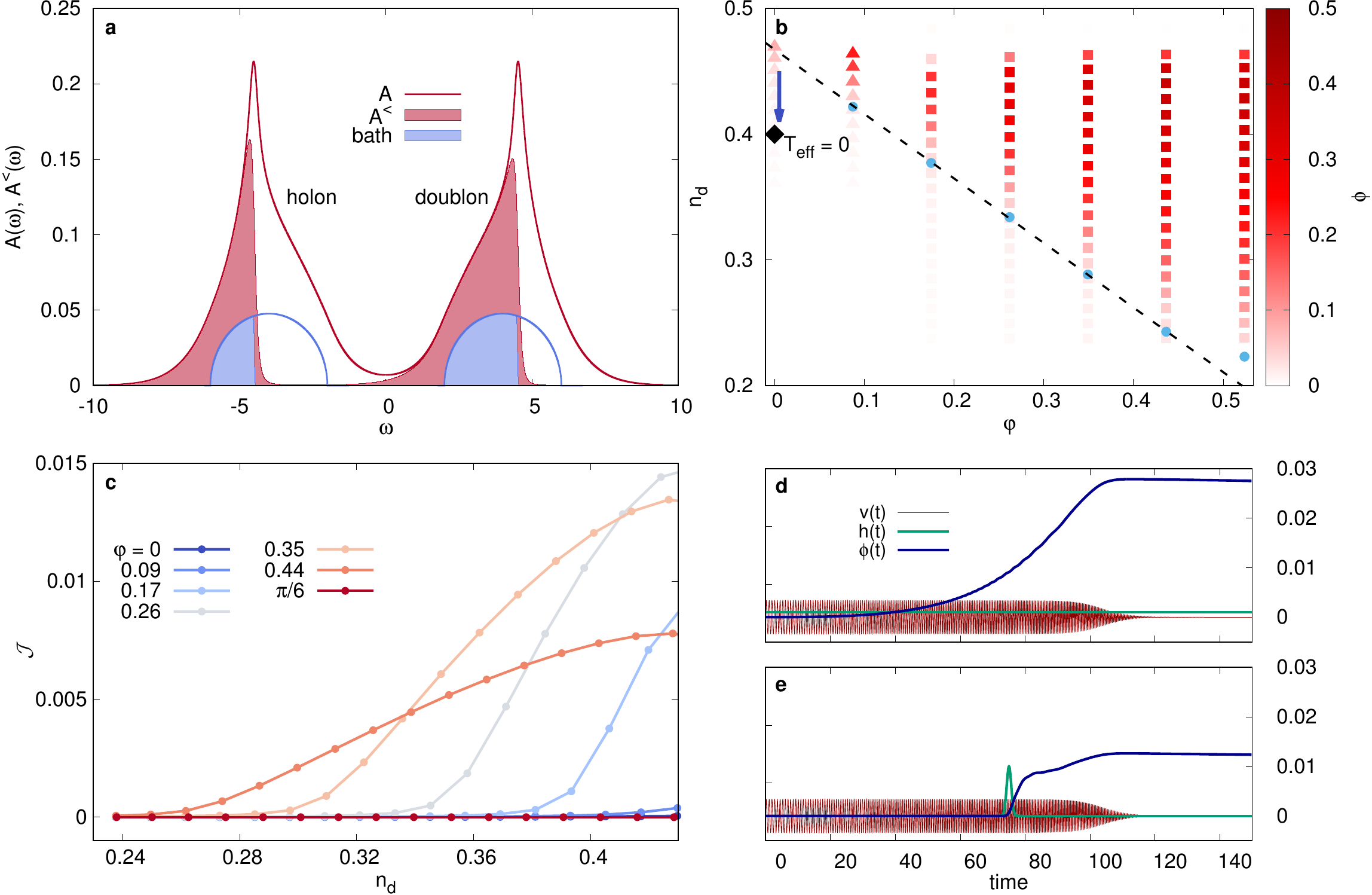}
\caption{{\bf The chiral $120^\circ$ condensate on the Bethe lattice.} In panels (a,b,c) the system is driven by two fermion baths, where the ``upper" bath injects high-energy electrons to form doublons, while the ``lower" bath absorbs low-energy electrons on singly occupied sites to form holons. Panels (d,e) show real-time simulations with an explicit photo-doping pulse. (a) Spectral function and occupation at $\varphi=0$ for the driven Bethe lattice, with $\mu_b=4.5, T_b=0.01,\Gamma=0.045,U=8.0$ and bandwidth $4$ (in units of $t_0$). The red-shaded area indicates the occupation of the Hubbard bands, while the blue-shaded area indicates the bath occupation. (b) The $n_d$--$\varphi$ phase diagram for the $120^\circ$ condensate sampled by varying the parameters of the fermion baths. The phase $\varphi$ is in radians and varies from $0$ to $\pi/6$. $W=2$ for the square symbols and $W=2.7$ for the triangles. The blue dots are obtained by fitting the phase boundary.  The black diamond is the extrapolated critical $n_d$ for $T_{\rm eff}=0$ and $\varphi=0$.  (c) The persistent current evaluated with $T_b=0.01,W=2$. A weak symmetry-breaking field $h=0.001$ is used in the DMFT iterations.  (d) and (e) Transient $120^\circ$-ordered states obtained by means of the entropy cooling protocol for the Bethe lattice with $U=9.0$ and $\varphi=0$ using two ways to break the symmetry  by a small pair field (green curves $h(t)$). $v(t)$ is the oscillating system-bath coupling which produces the photo-doped state. It is plotted with arbitrary units. }
\label{spec}
\end{figure*}
\begin{figure*}[t]
\includegraphics[scale=0.7]{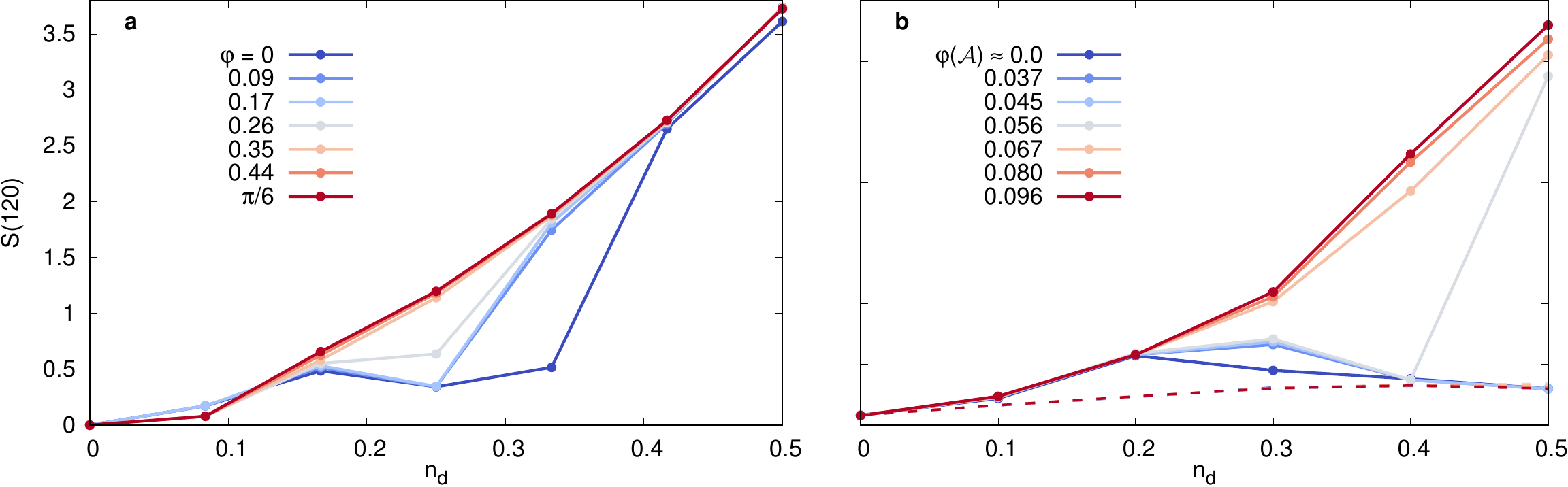}
\caption{{\bf Exact diagonalization results for the triangular and Kagome lattices.} (a) The pairing structure factor $S(120)$ of the $120^\circ$-twisted condensate versus the doublon number per site $n_d$ for the 12-site triangular lattice at $T_\text{eff}=0$. The Peierls phase $\varphi$ is varied from $0$ to $\pi/6$. (b) The pairing structure factor of the (${\bm q}=0$) $120^\circ$ order as a function of the doublon density for the 12-site Kagome lattice. The amplitude $\mathcal{A}$ of the dressing field is varied from $0$ to $1.4$ to produce different $\varphi$ values, as indicated in the label. Here, an external driving with frequency $\Omega= 5$ is assumed. 
The dashed line shows the structure factor for the uniform order. $J_{\perp}=-J_{z}=0.5$ is assumed and we impose a spatially homogeneous solution.}
\label{ed}
\end{figure*}

\section{\NoCaseChange{Doublon-holon condensate on three-colorable lattices}} 

We consider a half-filled fermionic Hubbard model under external driving, focusing on the triangular, the Kagome, and the Bethe lattice, see Fig.~\ref{diagram}. 
All these lattices are 3-colorable (we use the color labels $R$, $G$, $B$) and consist of connected $RGB$ chains or loops, i.e., triangular motifs.
The Hamiltonian reads 
\begin{align}
H=-t_0\sum_{\langle ij\rangle\sigma}e^{i\varphi_{ij}}c^\dag_{i\sigma}c_{j\sigma}+U\sum_{i} n_{i\uparrow}n_{i\downarrow} + g H_{\rm dr},
\label{model}
\end{align}
where $c_{i\sigma}$ is the electron annihilation operator at site $i$ with spin $\sigma$, $\langle ij\rangle$ denotes nearest-neighbor pairs, and $t_0$ and $U>0$ are hopping and interaction parameters. The Peierls phases $\varphi_{ij}$ represent an artificial gauge field implemented through a periodic (Floquet) modulation \cite{struck2012,hauke2012}. As will be demonstrated below, the gauge field can be used to tune the superconducting (SC) condensate's stability, but the twisted condensate also exists for $\varphi=0$ on the triangular and Bethe lattices.
To realize a long-lived photodoped state, we assume that a driving term $gH_{\rm dr}$ with an overall amplitude $g$ generates a nonthermal population of doublons and holons. One example of  
$H_{\rm dr}$ is a resonant optical excitation between the Hubbard bands, as widely adopted in experiments. Another example is the coupling of the system to two separate fermion baths (electrodes), as explained in Methods, and widely used in theoretical studies to emulate the photoexcitation protocol. 

We will focus on the strong interaction regime with a weak driving $g\ll t_0\ll U$ which is nearly resonant with the Mott gap. 
In this regime one generically finds a stationary nonequilibrium state whose properties are independent of the details of the driving. The system is Mott insulating in equilibrium, and the photodoped carriers have a long lifetime due to the large Mott gap \cite{sensarma2010,eckstein2011,mitrano2014}. An effective description of the photodoped state can be obtained from a $1/U$ expansion. As $g\ll t_0$, the driving term does not affect $H_{\rm eff}$ to leading order, but it can control the doublon density. 
Analogous to the doped Mott insulators at equilibrium, the effective physics of the photodoped state is well described by a generalized $t$-$J$ model \cite{li2020,kaneko2020,murakami2022} $H_{\rm eff}=H_t+H_J+H_{dh}$ with hopping $H_t=-t_0\sum_{\langle ij\rangle\sigma}e^{i\varphi_{ij}}[n_{i\bar{\sigma}}c^\dag_{i\sigma}c_{j\sigma}n_{j\bar{\sigma}}+\bar{n}_{i\bar{\sigma}}c^\dag_{i\sigma}c_{j\sigma}\bar{n}_{j\bar{\sigma}}]+\text{h.c.}$ and spin exchange $H_J=\sum_{\langle ij\rangle}J_{\rm ex}\boldsymbol{S}_i\cdot \boldsymbol{S}_j$, where $J_{\rm ex}=4t_0^2/U$. 
We have defined $\bar{n}_{i\bar{\sigma}}=1-n_{i,-\sigma}$. The doublon-holon interaction term reads
\begin{align}
H_{dh}=\frac{J_{\perp}}{2}\sum_{\langle ij\rangle}(e^{2i\varphi_{ij}}\phi^+_i\phi^-_j+\text{h.c.})+J_{z}\sum_{\langle ij\rangle}\phi^z_i\phi^z_j,
\label{Heff}
\end{align}
where the pairing operators $\phi^+_i=(\phi^-_i)^\dag=c^\dag_{i\uparrow}c^\dag_{i\downarrow}$ and $\phi^z_i=(n_i-1)/2$ span a pseudospin $\mathfrak{su}(2)$ algebra similar to that of spin $\bm S_i$. The original model \eqref{model} yields $J_{\perp}=-J_{z}=J_{\rm ex}$. 
 The first term in Eq.~\eqref{Heff}
originates from a doublon-holon exchange process, illustrated in Fig.~\ref{diagram}(a), which favors a doublon-holon condensation with $\langle \phi^+_i\rangle\ne0$. In solids, the second term is generically renormalized by the intersite Coulomb repulsion which suppresses charge segregation. We will focus in the following on uniform phases with $\langle n_i\rangle=1$ ($\langle \phi^z_i\rangle=0$). 

In pump-probe experiments on Mott insulating solids, a strong pump pulse is often applied for a short duration to create a quasi-stationary photodoped state, which can also be described by the above Hamiltonian $H_{\rm eff}$ with $g=0$. In either set-up, the effective theory for the nonequilibrium state of the Hubbard model is a generalized $t$-$J$ model of $n_d$ doublons and holons and $n_s=1-2n_d$ unpaired electrons per site. 

In this prethermal phase, the positive exchange amplitude $J_{\perp}$ tends to impose a staggered phase twist for the doublon-holon condensate ($\eta$--pairing) \cite{yang1989}, but this alternating SC order is generically impossible on a frustrated non-bipartite lattice. Instead, we consider a $120^\circ$ twisted pairing for the three-colorable lattices, defined by $\langle\phi^+_{i \in R}\rangle=e^{- i2\pi/3}\langle\phi^+_{i \in G}\rangle=e^{- i4\pi/3}\langle\phi^+_{i \in B}\rangle=\phi_0$, which spontaneously breaks time-reversal and inversion symmetry. To understand the energetics of the condensate, we can examine the mean-field energy for the order $\langle\phi^+_i\rangle=\phi_0e^{i\bm q\cdot\bm r_i}$ with momentum $\bm q$, given by $\langle H_{dh}\rangle/N_{\rm site}=|\phi_0|^2\epsilon({\bm q})$ per site, see Supplemental Note 1. In the following, we restrict ourselves to the case $\varphi_{ij}=\varphi$ along each bond of an $R\to G\to B$ cycle. For the triangular lattice, the above $120^\circ$ order is of momentum $\bm q =\frac{2\pi}{3}\bm b_1-\frac{2\pi}{3}\bm b_2$ with reciprocal lattice vectors $\bm b_1,\bm b_2$, and corresponds to one of the two chiral minima of the energy dispersion. This minimum can be further 
stabilized 
by an artificial Peierls phase $0<\varphi<\pi/3$, with $\varphi=\pi/6$ realizing the most stable condensate. The above discussion applies equally to the opposite chirality with a reversed phase twist. The Kagome lattice has three sites in a unit cell, giving rise to three bands in $\epsilon(\bm q)$. For $J_{\perp}>0$ the lowest-lying band for $J_{\perp}>0$ is flat for $\varphi=0$, which implies  
that no ordering pattern is singled out as energetically most favorable.
The artificial gauge field $\varphi_{ij}$ can however distort the flat band and favor a certain $120^\circ$ twisted order. As we shall show later in the paper, the order with ${\bm q}=0$ illustrated in Fig.~\ref{diagram}(c) can be stabilized by an artificial gauge field generated by circularly polarized light. 

It remains to be shown that the $120^\circ$ condensate is stable against quantum fluctuations. We first consider the maximum photodoping situation ($n_d=0.5$) where all sites are either doubly occupied or empty. In this case $H_{\rm eff}=H_{dh}$ corresponds to an XXZ model of pseudospin ${\bm \phi}$. With spatial homogeneity assumed, it is known that the $120^\circ$ condensate is generically stabilized on the triangular lattice. Furthermore, the $120^\circ$ condensates constitute the exact ground-state manifold if $J_{z}=J_{\perp}\cos(2\pi/3+\varphi)$ ($J_{z}=-J_{\perp}/2$ for $\varphi=0$, which may be realized with an NN Coulomb repulsion) for both the triangular and Kagome lattices considered here, see Supplemental Note 1 for a proof following the idea of Ref.~\citenum{changlani2018}. Away from maximum photodoping, the twisted $120^\circ$ condensate is challenged by the presence of singly occupied sites, and  
the additional 
terms in $H_{\rm eff}$, such as electron hopping, while the condensate should survive at least for $1/2-n_d\ll \mathcal{O}(t_0/U)$. We will numerically confirm that this condensate is in fact stable in an extended parameter regime away from the maximum photodoping limit $n_d=1/2$.

\section{\NoCaseChange{Optical response of the $120^\circ$ chiral condensate}} 
The twisted $120^\circ$ condensate embodies a spatially varying phase twist
and thus carries a persistent current even in the absence of an external field. With an external vector potential $A_{ij}$ along bond $\langle ij\rangle$, the doublon-holon current contribution along the cycle $R\to G\to B$ is 
$\mathcal{J}^{dh}_{ij}(A_{ij})=\delta H_{dh}/\delta A_{ij}\approx2\mathcal{J}^{dh}_0\sin(2\pi/3+2\varphi+2A_{ij})$ on the mean-field level, where $\mathcal{J}^{dh}_0=-2eJ_{1}|\phi_0|^2$ with the elementary charge $e$. A persistent current $\mathcal{J}^{dh}_{ij}(0)=2\mathcal{J}^{dh}_0\sin(2\pi/3+2\varphi)$ flows even when $A_{ij}=0$, see arrows in Fig.~\ref{diagram}. Indeed, the phase-twisted condensate can be viewed as a frustrated array of Josephson junctions \cite{theron1994}. 

As usual, a macroscopic superconducting current emerges when the condensate undergoes a uniform electric pulse $\bm E(t)$, which generates a vector potential $\bm A(t)=-\int^t ds\bm E(s)$. However, the breaking of the time-reversal and inversion symmetries allows for a nonlinear and anisotropic supercurrent response, of the general form
$\mathcal{J}^a= D^{ab}A_b+T^{abc}A_b A_c+\cdots$ with indices $a,b,c=x/y$ and Einstein convention. This is in contrast to conventional superconductors, where the fully symmetric tensor $T^{abc}$ vanishes due to unbroken inversion or time-reversal symmetries, which 
imply
$\bm J(-\bm A)=-\bm J(\bm A)$, and where the London equation $\bm J\propto \bm A$ usually holds. 

In particular, the chiral condensate allows for a nonlinear transverse current response perpendicular to $\bm A$, which constitutes a characteristic signature of the symmetry breaking. For the triangular lattice in Fig.~\ref{diagram}(b), the three-fold dihedral symmetry ($D_3$) imposes $T^{xyy}=T^{xxx}=0$, but allows for nonzero entries obeying $T^{xxy}=-T^{yyy}$.  The value of $T^{xxy}$ is determined by evaluating the gauge-invariant supercurrent density $\mathcal{J}^a=\frac{1}{S}\langle\frac{\delta H_{dh}}{\delta{A}_a}\rangle$ with $a=x,y$ and the total area $S$. At the mean-field level, one obtains $D^{ab}=\frac{\phi_0^2}{S_{\rm u.c.}}\frac{\partial^2 \epsilon(\bm q+\bm A)}{\partial A_a\partial A_b}$ and $T^{abc}=\frac{\phi_0^2}{2S_{\rm u.c.}}\frac{\partial^3 \epsilon(\bm q+\bm A)}{\partial A_a\partial A_b\partial A_c}$, with the unit-cell area $S_{\rm u.c.}$, see Supplemental Note 2. Here we consider $\bm A$ along the $x$--direction. The second-order response emerges due to the trigonal warping $\partial^3\epsilon/\partial k_x\partial k_x\partial k_y$ of the energy dispersion near the condensation minimum.

The current density is illustrated in Fig.~\ref{diagram}(d) and depends strongly on the artificial gauge field $\varphi$. In particular, the transverse current vanishes for $\varphi=\pi/6$. 
The nonlinear and anisotropic response
is phenomenologically similar to the second-order anomalous Hall effect \cite{sodemann2015,nagaosa2017} with inversion symmetry breaking, but in contrast to these works, it appears in the context of a SC response and requires the breaking of time-reversal symmetry. It is intriguing to note that if one applies a continuous-wave sinusoidal driving along $x$ ($E_x(t)= E_0\sin(\omega t)$), the transverse current oscillates at frequency $2\omega$. This Hall-like response and second-harmonic generation could serve as a smoking gun of the chiral order and can be tested with a four-point measurement.

\section{\NoCaseChange{Numerical determination of the phase diagram}} 
In the remainder of the article, we will explore the phase diagram of the chiral condensate on three different frustrated lattices: the Bethe, the triangular, and the Kagome lattices. To provide a concrete example of the $120^\circ$ twisted order, we first consider a numerically solvable model in the thermodynamic limit, the driven Hubbard model on the Bethe lattice with hopping $t_0/\sqrt{z}$ where the coordination number is taken to $z\to\infty$. In this case, $t_0$ is taken as the unit of energy. While this lattice has no loops, the driven system can support the twisted condensate and, most importantly, 
 can be 
solved using nonequilibrium dynamical mean-field theory (DMFT) in the strong interaction regime \cite{georges1996, aoki2014}.   
We drive the system by coupling it to two fermion baths with a semielliptic density of states with temperature $T_b$ and different chemical potentials $\pm \mu_b$. As discussed above, the details of the driving do not matter as long as $g\ll t_0$. The baths can be exactly integrated out and incorporated into the DMFT iterations through a hybridization density of states $D_{\pm}(\epsilon)=\Gamma\sqrt{1-(\epsilon\pm U/2)^2/W^2}$ with $\Gamma = g^2/W$ and half-bandwidth $W$. The parameters $\pm \mu_b$ and $T_b$ are varied to control the doublon and holon distribution in the Hubbard system \cite{li2021}.  

We show the spectral function $A=-\frac{1}{\pi}\operatorname{Im}G^r$ and the occupation $A^<=\frac{1}{2\pi}\operatorname{Im}G^<$ for a typical driven state in Fig.~\ref{spec}(a). The external driving creates two separate Fermi surfaces in the lower and upper Hubbard bands around $\omega=\pm\mu_b$, indicating the presence of excess doublons and holons. The two Fermi energies correspond to the energy cost or gain associated with the replacement of a doublon or holon by an unpaired electron. Even though the system is highly excited, the Fermi edges can be rather sharp, with low effective temperatures $T_{\rm eff}$ defined by fitting the distribution function $f(\omega)=A^<(\omega)/A(\omega)$ separately near the two Fermi levels. This confirms that the system can be described by a constrained quasi-equilibrium with excess doublons and holons of density $n_d$ and temperature $T_{\rm eff}$, as already shown in previous works \cite{li2020,li2021,murakami2022}. The order parameter $\phi=\langle\phi^+\rangle$ is sampled for various bath parameters, which allows to generate the phase diagram for the $120^\circ$ condensate shown in Fig.~\ref{spec}(b). The system exhibits a transition to the $120^\circ$-ordered phase beyond a critical double occupancy $n^c_d$, indicated by blue dots (see Supplemental Note 3), and is enhanced as $\varphi$ increases. A persistent current $\mathcal{J}$ flows along $R\to G\to B$, whose magnitude is plotted in panel (c). In experiments, carefully designed protocols can minimze the entropy production and thus achieve low effective temperatures \cite{werner2019}. In our calculations, we can tune the effective temperature of the doublons and holons by changing bath parameters, and then extrapolate the phase boundary to $T_{\rm eff}=0$. This procedure yields the upper bound of the critical doublon number $n^c_d\approx 0.4$ at $\varphi=0$ (black diamond in (b)).  

The above results for the Bethe lattice confirm the existence of the $120^\circ$ SC condensate in a wide parameter range, which extends to $\varphi=0$ and at least down to $n_d\approx 0.4$, and also the validity of the effective theory~\eqref{Heff} at finite $U$. It remains to be investigated if this intriguing state can be realized on an ultrafast timescale. To address this question, we consider a real-time entropy-cooling protocol \cite{werner2019,werner2019prb}, which allows to generate cold photodoped states. Here, the system is coupled to two narrow bands with width $W=0.1$ centered at $\omega=\pm6.0$ and identical chemical potentials $\mu_b=0$, while the system-bath coupling $v(t)$ oscillates fast to excite photocarriers, see Methods.  
As shown in Fig.~\ref{spec}(d), a nonzero $120^\circ$ SC order parameter quickly emerges at $\varphi=0$ under the driving $v(t)$, in the presence of a small symmetry-breaking field $h=0.001$ coupling to $\phi^+_i$. In the case of transition metal compounds, our unit of time is of the order of femtoseconds. Within the time range accessible in our numerical simulations, the order appears to decay very slowly after the driving is switched off. In addition, we have also simulated a system perturbed by a short pulse of the symmetry-breaking field $h(t)$, see Fig.~\ref{spec}(e). In this case, the $120^\circ$ order continues to grow after the pulse, which strongly suggests a spontaneous symmetry breaking. 

We now use exact diagonalization to directly treat the generalized $t$-$J$ model \eqref{Heff} on a triangular lattice, to address the existence of the condensate in 2D systems at $T_{\rm eff}=0$. We study an $N_{\rm site}=12$ cluster with periodic boundary conditions, and calculate the pairing structure factor at the $120^\circ$ point, namely $S(120)=\sum_{ij}\theta_i\theta^*_j\langle \phi^+_i\phi^-_j\rangle/N_{\rm site}$ with $\theta_{i\in R}=e^{i2\pi/3}\theta_{i\in G}=e^{i4\pi/3}\theta_{i\in B}=1$ corresponding to the ``color". $S(120)$ quantifies the total $120^\circ$ order. The result is shown in Fig.~\ref{ed}(a), and 
suggests 
the stability of the $120^\circ$ condensate away from $n_d=1/2$ down to $\varphi=0$, in line with the observation for the Bethe lattice from DMFT. The critical doublon number is $n^c_d(\varphi=0)\sim 0.37$.

\section{\NoCaseChange{Realization on the Kagome lattice}} 
Finally, we comment on how to realize the $120^\circ$ SC order on the Kagome lattice. The generalized $t$-$J$ model in the maximum photodoping limit ($n_d=0.5$) is equivalent to an $XXZ$ spin model, which exhibits a complex phase diagram on the Kagome lattice. As discussed above, the key task is to induce an artificial gauge field which favors a specific $120^\circ$ ordering pattern, 
for example the $\bm q =0$ three-coloring of the lattice shown
in Fig.~\ref{diagram}(c). In the case of solid-state systems, we consider applying a strong laser with circular polarization, similar to the setup in Ref.~\cite{claassen2017}, with a vector potential $\bm A(t)=\mathcal{A}_0(\cos(\Omega t), -\sin(\Omega t))$.
The field couples through a time-dependent Peierls phase $\exp(i(\bm r_{i}-\bm r_j)\cdot \bm A(t))$ to the bond $\langle ij\rangle$. In the high-frequency limit ($\Omega\gg t_0$, off-resonant with $U$), the hopping parameters are renormalized, leading to a complex nearest-neighbor (NN) hopping $t_{\rm R}=|t_{\rm R}|e^{i\varphi(\mathcal{A})}$. 
This renormalized hopping appears to favor the ${\bm q} = 0$ condensate shown in Fig.~\ref{diagram}(c). The next-nearest neighbor hopping which is also created and would favor a uniform order is usually very small, see the Supplemental Note 4.

We study the resulting effective model on a 12-site cluster with exact diagonalization, 
defining $\mathcal{A}=\mathcal{A}_0 a/2$ with lattice constant $a$. The results are shown in Fig.~\ref{ed}(b), and clearly indicate the appearance of the (${\bm q} = 0$) $120^\circ$ SC condensate for $\varphi(\mathcal{A})\gtrsim0.067$ and $n_d\gtrsim 0.3$. The dependence of $S(120)$ on $n_d$ is qualitatively similar for both the triangular and Kagome lattices above the transition. In contrast, the uniform order (indicated by the dashed line) is always suppressed. It is worth noting that other orders, such as an $\sqrt{3}\times\sqrt{3}$ type $120^\circ$ order cannot be studied with the small cluster shown in Fig.~\ref{diagram}.  
The 
two orders share the identical mean-field energy at $\varphi=0$. However, a nonzero $|\varphi|<\pi/3$ stabilizes the ${\bm q} = 0$ order considered here, see Supplemental Note 4. The $\sqrt{3}\times\sqrt{3}$ type order can however be stabilized with a different gauge field, see Supplemental Note 4. Finally we note that other terms, which are generated by the Floquet driving but ignored here, can alter the phase diagram at the quantitative level.

\section{\NoCaseChange{Conclusion}} 
Our work established a new type of chiral superconductivity in photodoped Mott insulators, which breaks the time-reversal and inversion symmetries through a spatially twisted order parameter, namely the $120^\circ$ condensate. This condensate originates from a positive doublon-holon exchange amplitude $J_{\perp}>0$, which is intrinsically related to the nonequilibrium nature of the photodoped states and contrasts with the negative exchange in equilibrium BEC-like pairing induced by charge attraction \cite{micnas1990}. The exchange processes furthermore generate a doublon-holon interaction (the $J_{z}$ term), which favors charge segregation. This effect is ignored here, since it should be suppressed by the inter-site Coulomb repulsion in solids. Also, if the ordered phase is created by an ultrafast uniform excitation, we can assume that the state remains homogeneous on the femtoseconds timescale. In the presence of a light-induced artificial gauge field, the order can be further enhanced and even stabilized on the Kagome lattice. The persistent loop current and the nonlinear transverse superconducting current are characteristic signatures of the chiral $120^\circ$ condensate and allow to realize a second harmonic generation. The phenomenology contrasts with the conventional description of the SC electromagnetic response based on the linear London equation $\bm j\propto \bm A$, where even-order responses are excluded by time-reversal and inversion symmetry. 

In experiments, the photodoped state can be realized by applying femtosecond laser pulses to condensed matter systems \cite{stojchevska2014}, or via the 
tilting of 
optical lattices \cite{greif2011}. When low entropies are maintained, both protocols allow to create long-lived photodoped doublons and holons in the presence of a large Mott gap \cite{sensarma2010,eckstein2011,mitrano2014}, which quickly relax to a prethermal regime characterized by the generalized $t$-$J$ physics. Since the condensate exists down to $\varphi=0$, it can be relevant to the photodoped SC states in correlated materials of triangular lattice geometry, such as $\kappa$--(BEDT--TTF)$_2$Cu[N(CN)$_2$]Br \cite{buzzi2020}. 
Finally, doped Mott insulators have a very rich phase diagram, and the competition between the chiral condensate and other long-lived and hidden phases, especially away from maximum photodoping, is an interesting topic for further investigations. 

\bibliographystyle{naturemag}
\bibliography{ref}

\begin{thebibliography}{10}
\expandafter\ifx\csname url\endcsname\relax
  \def\url#1{\texttt{#1}}\fi
\expandafter\ifx\csname urlprefix\endcsname\relax\def\urlprefix{URL }\fi
\providecommand{\bibinfo}[2]{#2}
\providecommand{\eprint}[2][]{\url{#2}}

\bibitem{kiefl1990}
\bibinfo{author}{Kiefl, R.~F.} \emph{et~al.}
\newblock \bibinfo{title}{Search for anomalous internal magnetic fields in
  high-${{{\mathit{T}}_{\mathit{c}}}}$ superconductors as evidence for broken
  time-reversal symmetry}.
\newblock \emph{\bibinfo{journal}{Phys. Rev. Lett.}}
  \textbf{\bibinfo{volume}{64}}, \bibinfo{pages}{2082--2085}
  (\bibinfo{year}{1990}).
\newblock \urlprefix\url{https://link.aps.org/doi/10.1103/PhysRevLett.64.2082}.

\bibitem{mackenzie2003}
\bibinfo{author}{Mackenzie, A.~P.} \& \bibinfo{author}{Maeno, Y.}
\newblock \bibinfo{title}{The superconductivity of
  ${{{\mathrm{Sr}}_{2}{\mathrm{RuO}}_{4}}}$ and the physics of spin-triplet
  pairing}.
\newblock \emph{\bibinfo{journal}{Rev. Mod. Phys.}}
  \textbf{\bibinfo{volume}{75}}, \bibinfo{pages}{657--712}
  (\bibinfo{year}{2003}).
\newblock \urlprefix\url{https://link.aps.org/doi/10.1103/RevModPhys.75.657}.

\bibitem{joynt2002}
\bibinfo{author}{Joynt, R.} \& \bibinfo{author}{Taillefer, L.}
\newblock \bibinfo{title}{The superconducting phases of
  ${{\mathrm{UPt}}_{3}}$}.
\newblock \emph{\bibinfo{journal}{Rev. Mod. Phys.}}
  \textbf{\bibinfo{volume}{74}}, \bibinfo{pages}{235--294}
  (\bibinfo{year}{2002}).
\newblock \urlprefix\url{https://link.aps.org/doi/10.1103/RevModPhys.74.235}.

\bibitem{avers2020}
\bibinfo{author}{Avers, K.~E.} \emph{et~al.}
\newblock \bibinfo{title}{Broken time-reversal symmetry in the topological
  superconductor {UPt$_3$}}.
\newblock \emph{\bibinfo{journal}{Nat. Phys.}} \textbf{\bibinfo{volume}{16}},
  \bibinfo{pages}{531--535} (\bibinfo{year}{2020}).
\newblock \urlprefix\url{https://doi.org/10.1038/s41567-020-0822-z}.

\bibitem{cao2016}
\bibinfo{author}{Cao, Y.} \emph{et~al.}
\newblock \bibinfo{title}{Superlattice-induced insulating states and
  valley-protected orbits in twisted bilayer graphene}.
\newblock \emph{\bibinfo{journal}{Phys. Rev. Lett.}}
  \textbf{\bibinfo{volume}{117}}, \bibinfo{pages}{116804}
  (\bibinfo{year}{2016}).
\newblock
  \urlprefix\url{https://link.aps.org/doi/10.1103/PhysRevLett.117.116804}.

\bibitem{ichikawa2011}
\bibinfo{author}{Ichikawa, H.} \emph{et~al.}
\newblock \bibinfo{title}{Transient photoinduced `hidden' phase in a
  manganite}.
\newblock \emph{\bibinfo{journal}{Nat. Mater.}} \textbf{\bibinfo{volume}{10}},
  \bibinfo{pages}{101} (\bibinfo{year}{2011}).
\newblock \urlprefix\url{https://doi.org/10.1038/nmat2929}.

\bibitem{stojchevska2014}
\bibinfo{author}{Stojchevska, L.} \emph{et~al.}
\newblock \bibinfo{title}{Ultrafast switching to a stable hidden quantum state
  in an electronic crystal}.
\newblock \emph{\bibinfo{journal}{Science}} \textbf{\bibinfo{volume}{344}},
  \bibinfo{pages}{177--180} (\bibinfo{year}{2014}).
\newblock \urlprefix\url{https://doi.org/10.1126/science.1241591}.

\bibitem{depaz2013}
\bibinfo{author}{de~Paz, A.} \emph{et~al.}
\newblock \bibinfo{title}{Nonequilibrium quantum magnetism in a dipolar lattice
  gas}.
\newblock \emph{\bibinfo{journal}{Phys. Rev. Lett.}}
  \textbf{\bibinfo{volume}{111}}, \bibinfo{pages}{185305}
  (\bibinfo{year}{2013}).
\newblock
  \urlprefix\url{https://link.aps.org/doi/10.1103/PhysRevLett.111.185305}.

\bibitem{singh2019}
\bibinfo{author}{Singh, K.} \emph{et~al.}
\newblock \bibinfo{title}{Quantifying and controlling prethermal nonergodicity
  in interacting floquet matter}.
\newblock \emph{\bibinfo{journal}{Phys. Rev. X}} \textbf{\bibinfo{volume}{9}},
  \bibinfo{pages}{041021} (\bibinfo{year}{2019}).
\newblock \urlprefix\url{https://link.aps.org/doi/10.1103/PhysRevX.9.041021}.

\bibitem{fausti2011}
\bibinfo{author}{Fausti, D.} \emph{et~al.}
\newblock \bibinfo{title}{Light-induced superconductivity in a stripe-ordered
  cuprate}.
\newblock \emph{\bibinfo{journal}{Science}} \textbf{\bibinfo{volume}{331}},
  \bibinfo{pages}{189--191} (\bibinfo{year}{2011}).
\newblock \urlprefix\url{https://science.sciencemag.org/content/331/6014/189}.

\bibitem{mitrano2016}
\bibinfo{author}{Mitrano, M.} \emph{et~al.}
\newblock \bibinfo{title}{Possible light-induced superconductivity in
  {K$_3$C$_{60}$} at high temperature}.
\newblock \emph{\bibinfo{journal}{Nature}} \textbf{\bibinfo{volume}{530}},
  \bibinfo{pages}{461} (\bibinfo{year}{2016}).
\newblock \urlprefix\url{https://doi.org/10.1038/nature16522}.

\bibitem{buzzi2020}
\bibinfo{author}{Buzzi, M.} \emph{et~al.}
\newblock \bibinfo{title}{Photomolecular high-temperature superconductivity}.
\newblock \emph{\bibinfo{journal}{Phys. Rev. X}} \textbf{\bibinfo{volume}{10}},
  \bibinfo{pages}{031028} (\bibinfo{year}{2020}).
\newblock \urlprefix\url{https://link.aps.org/doi/10.1103/PhysRevX.10.031028}.

\bibitem{budden2021}
\bibinfo{author}{Budden, M.} \emph{et~al.}
\newblock \bibinfo{title}{Evidence for metastable photo-induced
  superconductivity in {K$_3$C$_60$}}.
\newblock \emph{\bibinfo{journal}{Nat. Phys.}} \textbf{\bibinfo{volume}{17}},
  \bibinfo{pages}{611--618} (\bibinfo{year}{2021}).
\newblock \urlprefix\url{https://doi.org/10.1038/s41567-020-01148-1}.

\bibitem{iwai2003}
\bibinfo{author}{Iwai, S.} \emph{et~al.}
\newblock \bibinfo{title}{Ultrafast optical switching to a metallic state by
  photoinduced {Mott} transition in a {Halogen-Bridged Nickel}-chain compound}.
\newblock \emph{\bibinfo{journal}{Phys. Rev. Lett.}}
  \textbf{\bibinfo{volume}{91}}, \bibinfo{pages}{057401}
  (\bibinfo{year}{2003}).
\newblock
  \urlprefix\url{https://link.aps.org/doi/10.1103/PhysRevLett.91.057401}.

\bibitem{okamoto2010}
\bibinfo{author}{Okamoto, H.} \emph{et~al.}
\newblock \bibinfo{title}{Ultrafast charge dynamics in photoexcited
  ${{\text{Nd}}_{2}{\text{CuO}}_{4}}$ and ${{\text{La}}_{2}{\text{CuO}}_{4}}$
  cuprate compounds investigated by femtosecond absorption spectroscopy}.
\newblock \emph{\bibinfo{journal}{Phys. Rev. B}} \textbf{\bibinfo{volume}{82}},
  \bibinfo{pages}{060513} (\bibinfo{year}{2010}).
\newblock \urlprefix\url{https://link.aps.org/doi/10.1103/PhysRevB.82.060513}.

\bibitem{beaud2014}
\bibinfo{author}{Beaud, P.} \emph{et~al.}
\newblock \bibinfo{title}{A time-dependent order parameter for ultrafast
  photoinduced phase transitions}.
\newblock \emph{\bibinfo{journal}{Nat. Mater.}} \textbf{\bibinfo{volume}{13}},
  \bibinfo{pages}{923} (\bibinfo{year}{2014}).
\newblock \urlprefix\url{https://doi.org/10.1038/nmat4046}.

\bibitem{dean2016}
\bibinfo{author}{Dean, M.} \emph{et~al.}
\newblock \bibinfo{title}{Ultrafast energy-and momentum-resolved dynamics of
  magnetic correlations in the photo-doped {Mott} insulator {Sr$_2$IrO$_4$}}.
\newblock \emph{\bibinfo{journal}{Nat. Mater.}} \textbf{\bibinfo{volume}{15}},
  \bibinfo{pages}{601} (\bibinfo{year}{2016}).
\newblock \urlprefix\url{https://doi.org/10.1038/nmat4641}.

\bibitem{ligges2018}
\bibinfo{author}{Ligges, M.} \emph{et~al.}
\newblock \bibinfo{title}{Ultrafast doublon dynamics in photoexcited
  ${1T}$-${{\mathrm{TaS}}_{2}}$}.
\newblock \emph{\bibinfo{journal}{Phys. Rev. Lett.}}
  \textbf{\bibinfo{volume}{120}}, \bibinfo{pages}{166401}
  (\bibinfo{year}{2018}).
\newblock
  \urlprefix\url{https://link.aps.org/doi/10.1103/PhysRevLett.120.166401}.

\bibitem{rosch2008}
\bibinfo{author}{Rosch, A.}, \bibinfo{author}{Rasch, D.},
  \bibinfo{author}{Binz, B.} \& \bibinfo{author}{Vojta, M.}
\newblock \bibinfo{title}{Metastable superfluidity of repulsive fermionic atoms
  in optical lattices}.
\newblock \emph{\bibinfo{journal}{Phys. Rev. Lett.}}
  \textbf{\bibinfo{volume}{101}}, \bibinfo{pages}{265301}
  (\bibinfo{year}{2008}).
\newblock
  \urlprefix\url{https://link.aps.org/doi/10.1103/PhysRevLett.101.265301}.

\bibitem{peronaci2020}
\bibinfo{author}{Peronaci, F.}, \bibinfo{author}{Parcollet, O.} \&
  \bibinfo{author}{Schir\'o, M.}
\newblock \bibinfo{title}{Enhancement of local pairing correlations in
  periodically driven {{Mott}} insulators}.
\newblock \emph{\bibinfo{journal}{Phys. Rev. B}}
  \textbf{\bibinfo{volume}{101}}, \bibinfo{pages}{161101}
  (\bibinfo{year}{2020}).
\newblock \urlprefix\url{https://link.aps.org/doi/10.1103/PhysRevB.101.161101}.

\bibitem{li2020}
\bibinfo{author}{Li, J.}, \bibinfo{author}{Golez, D.}, \bibinfo{author}{Werner,
  P.} \& \bibinfo{author}{Eckstein, M.}
\newblock \bibinfo{title}{${\ensuremath{\eta}}$-paired superconducting hidden
  phase in photodoped {Mott} insulators}.
\newblock \emph{\bibinfo{journal}{Phys. Rev. B}}
  \textbf{\bibinfo{volume}{102}}, \bibinfo{pages}{165136}
  (\bibinfo{year}{2020}).
\newblock \urlprefix\url{https://link.aps.org/doi/10.1103/PhysRevB.102.165136}.

\bibitem{murakami2022}
\bibinfo{author}{Murakami, Y.} \emph{et~al.}
\newblock \bibinfo{title}{{Exploring nonequilibrium phases of photo-doped
  {Mott} insulators with generalized Gibbs ensembles}}.
\newblock \emph{\bibinfo{journal}{Commun. Phys.}} \textbf{\bibinfo{volume}{5}},
  \bibinfo{pages}{1--8} (\bibinfo{year}{2022}).
\newblock \urlprefix\url{https://doi.org/10.1038/s42005-021-00799-7}.

\bibitem{kalmeyer1987}
\bibinfo{author}{Kalmeyer, V.} \& \bibinfo{author}{Laughlin, R.~B.}
\newblock \bibinfo{title}{Equivalence of the resonating-valence-bond and
  fractional quantum {Hall} states}.
\newblock \emph{\bibinfo{journal}{Phys. Rev. Lett.}}
  \textbf{\bibinfo{volume}{59}}, \bibinfo{pages}{2095--2098}
  (\bibinfo{year}{1987}).
\newblock \urlprefix\url{https://link.aps.org/doi/10.1103/PhysRevLett.59.2095}.

\bibitem{struck2013}
\bibinfo{author}{Struck, J.} \emph{et~al.}
\newblock \bibinfo{title}{{Engineering Ising-XY spin-models in a triangular
  lattice using tunable artificial gauge fields}}.
\newblock \emph{\bibinfo{journal}{Nat. Phys.}} \textbf{\bibinfo{volume}{9}},
  \bibinfo{pages}{738--743} (\bibinfo{year}{2013}).
\newblock \urlprefix\url{https://doi.org/10.1038/nphys2750}.

\bibitem{struck2012}
\bibinfo{author}{Struck, J.} \emph{et~al.}
\newblock \bibinfo{title}{Tunable gauge potential for neutral and spinless
  particles in driven optical lattices}.
\newblock \emph{\bibinfo{journal}{Phys. Rev. Lett.}}
  \textbf{\bibinfo{volume}{108}}, \bibinfo{pages}{225304}
  (\bibinfo{year}{2012}).
\newblock
  \urlprefix\url{https://link.aps.org/doi/10.1103/PhysRevLett.108.225304}.

\bibitem{hauke2012}
\bibinfo{author}{Hauke, P.} \emph{et~al.}
\newblock \bibinfo{title}{Non-abelian gauge fields and topological insulators
  in shaken optical lattices}.
\newblock \emph{\bibinfo{journal}{Phys. Rev. Lett.}}
  \textbf{\bibinfo{volume}{109}}, \bibinfo{pages}{145301}
  (\bibinfo{year}{2012}).
\newblock
  \urlprefix\url{https://link.aps.org/doi/10.1103/PhysRevLett.109.145301}.

\bibitem{claassen2017}
\bibinfo{author}{Claassen, M.}, \bibinfo{author}{Jiang, H.-C.},
  \bibinfo{author}{Moritz, B.} \& \bibinfo{author}{Devereaux, T.~P.}
\newblock \bibinfo{title}{Dynamical time-reversal symmetry breaking and
  photo-induced chiral spin liquids in frustrated {Mott} insulators}.
\newblock \emph{\bibinfo{journal}{Nat. Commun.}} \textbf{\bibinfo{volume}{8}},
  \bibinfo{pages}{1--9} (\bibinfo{year}{2017}).
\newblock \urlprefix\url{https://doi.org/10.1038/s41467-017-00876-y}.

\bibitem{shores2005}
\bibinfo{author}{Shores, M.~P.}, \bibinfo{author}{Nytko, E.~A.},
  \bibinfo{author}{Bartlett, B.~M.} \& \bibinfo{author}{Nocera, D.~G.}
\newblock \bibinfo{title}{{A structurally perfect S= 1/2 Kagome
  antiferromagnet}}.
\newblock \emph{\bibinfo{journal}{J. Am. Chem. Soc.}}
  \textbf{\bibinfo{volume}{127}}, \bibinfo{pages}{13462--13463}
  (\bibinfo{year}{2005}).
\newblock \urlprefix\url{https://doi.org/10.1021/ja053891p}.

\bibitem{sensarma2010}
\bibinfo{author}{Sensarma, R.} \emph{et~al.}
\newblock \bibinfo{title}{Lifetime of double occupancies in the fermi-{Hubbard}
  model}.
\newblock \emph{\bibinfo{journal}{Phys. Rev. B}} \textbf{\bibinfo{volume}{82}},
  \bibinfo{pages}{224302} (\bibinfo{year}{2010}).
\newblock \urlprefix\url{https://link.aps.org/doi/10.1103/PhysRevB.82.224302}.

\bibitem{eckstein2011}
\bibinfo{author}{Eckstein, M.} \& \bibinfo{author}{Werner, P.}
\newblock \bibinfo{title}{Thermalization of a pump-excited {Mott} insulator}.
\newblock \emph{\bibinfo{journal}{Phys. Rev. B}} \textbf{\bibinfo{volume}{84}},
  \bibinfo{pages}{035122} (\bibinfo{year}{2011}).
\newblock \urlprefix\url{https://link.aps.org/doi/10.1103/PhysRevB.84.035122}.

\bibitem{mitrano2014}
\bibinfo{author}{Mitrano, M.} \emph{et~al.}
\newblock \bibinfo{title}{Pressure-dependent relaxation in the photoexcited
  mott insulator {ET-F$_2$TCNQ}: Influence of hopping and correlations on
  quasiparticle recombination rates}.
\newblock \emph{\bibinfo{journal}{Phys. Rev. Lett.}}
  \textbf{\bibinfo{volume}{112}}, \bibinfo{pages}{117801}
  (\bibinfo{year}{2014}).
\newblock
  \urlprefix\url{https://link.aps.org/doi/10.1103/PhysRevLett.112.117801}.

\bibitem{kaneko2020}
\bibinfo{author}{Kaneko, T.}, \bibinfo{author}{Yunoki, S.} \&
  \bibinfo{author}{Millis, A.~J.}
\newblock \bibinfo{title}{{Charge stiffness and long-range correlation in the
  optically induced ${\ensuremath{\eta}}$-pairing state of the one-dimensional
  {Hubbard} model}}.
\newblock \emph{\bibinfo{journal}{Phys. Rev. Research}}
  \textbf{\bibinfo{volume}{2}}, \bibinfo{pages}{032027} (\bibinfo{year}{2020}).
\newblock
  \urlprefix\url{https://link.aps.org/doi/10.1103/PhysRevResearch.2.032027}.

\bibitem{yang1989}
\bibinfo{author}{Yang, C.~N.}
\newblock \bibinfo{title}{\ensuremath{\eta} pairing and off-diagonal long-range
  order in a {Hubbard} model}.
\newblock \emph{\bibinfo{journal}{Phys. Rev. Lett.}}
  \textbf{\bibinfo{volume}{63}}, \bibinfo{pages}{2144--2147}
  (\bibinfo{year}{1989}).
\newblock \urlprefix\url{https://link.aps.org/doi/10.1103/PhysRevLett.63.2144}.

\bibitem{changlani2018}
\bibinfo{author}{Changlani, H.~J.}, \bibinfo{author}{Kochkov, D.},
  \bibinfo{author}{Kumar, K.}, \bibinfo{author}{Clark, B.~K.} \&
  \bibinfo{author}{Fradkin, E.}
\newblock \bibinfo{title}{Macroscopically degenerate exactly solvable point in
  the spin-${1/2}$ {{Kagome}} quantum antiferromagnet}.
\newblock \emph{\bibinfo{journal}{Phys. Rev. Lett.}}
  \textbf{\bibinfo{volume}{120}}, \bibinfo{pages}{117202}
  (\bibinfo{year}{2018}).
\newblock
  \urlprefix\url{https://link.aps.org/doi/10.1103/PhysRevLett.120.117202}.

\bibitem{theron1994}
\bibinfo{author}{Th\'eron, R.}, \bibinfo{author}{Korshunov, S.~E.},
  \bibinfo{author}{Simond, J.~B.}, \bibinfo{author}{Leemann, C.} \&
  \bibinfo{author}{Martinoli, P.}
\newblock \bibinfo{title}{Observation of domain-wall superlattice states in a
  frustrated triangular array of {Josephson} junctions}.
\newblock \emph{\bibinfo{journal}{Phys. Rev. Lett.}}
  \textbf{\bibinfo{volume}{72}}, \bibinfo{pages}{562--565}
  (\bibinfo{year}{1994}).
\newblock \urlprefix\url{https://link.aps.org/doi/10.1103/PhysRevLett.72.562}.

\bibitem{sodemann2015}
\bibinfo{author}{Sodemann, I.} \& \bibinfo{author}{Fu, L.}
\newblock \bibinfo{title}{Quantum nonlinear {Hall} effect induced by {Berry}
  curvature dipole in time-reversal invariant materials}.
\newblock \emph{\bibinfo{journal}{Phys. Rev. Lett.}}
  \textbf{\bibinfo{volume}{115}}, \bibinfo{pages}{216806}
  (\bibinfo{year}{2015}).
\newblock
  \urlprefix\url{https://link.aps.org/doi/10.1103/PhysRevLett.115.216806}.

\bibitem{nagaosa2017}
\bibinfo{author}{Nagaosa, N.} \& \bibinfo{author}{Morimoto, T.}
\newblock \bibinfo{title}{Concept of quantum geometry in optoelectronic
  processes in solids: Application to solar cells}.
\newblock \emph{\bibinfo{journal}{Adv. Mater.}} \textbf{\bibinfo{volume}{29}},
  \bibinfo{pages}{1603345} (\bibinfo{year}{2017}).
\newblock \urlprefix\url{https://doi.org/10.1002/adma.201603345}.

\bibitem{georges1996}
\bibinfo{author}{Georges, A.}, \bibinfo{author}{Kotliar, G.},
  \bibinfo{author}{Krauth, W.} \& \bibinfo{author}{Rozenberg, M.~J.}
\newblock \bibinfo{title}{Dynamical mean-field theory of strongly correlated
  fermion systems and the limit of infinite dimensions}.
\newblock \emph{\bibinfo{journal}{Rev. Mod. Phys.}}
  \textbf{\bibinfo{volume}{68}}, \bibinfo{pages}{13--125}
  (\bibinfo{year}{1996}).
\newblock \urlprefix\url{https://link.aps.org/doi/10.1103/RevModPhys.68.13}.

\bibitem{aoki2014}
\bibinfo{author}{Aoki, H.} \emph{et~al.}
\newblock \bibinfo{title}{Nonequilibrium dynamical mean-field theory and its
  applications}.
\newblock \emph{\bibinfo{journal}{Rev. Mod. Phys.}}
  \textbf{\bibinfo{volume}{86}}, \bibinfo{pages}{779--837}
  (\bibinfo{year}{2014}).
\newblock \urlprefix\url{https://link.aps.org/doi/10.1103/RevModPhys.86.779}.

\bibitem{li2021}
\bibinfo{author}{Li, J.} \& \bibinfo{author}{Eckstein, M.}
\newblock \bibinfo{title}{Nonequilibrium steady-state theory of photodoped
  {Mott} insulators}.
\newblock \emph{\bibinfo{journal}{Phys. Rev. B}}
  \textbf{\bibinfo{volume}{103}}, \bibinfo{pages}{045133}
  (\bibinfo{year}{2021}).
\newblock \urlprefix\url{https://link.aps.org/doi/10.1103/PhysRevB.103.045133}.

\bibitem{werner2019}
\bibinfo{author}{Werner, P.}, \bibinfo{author}{Eckstein, M.},
  \bibinfo{author}{M{\"u}ller, M.} \& \bibinfo{author}{Refael, G.}
\newblock \bibinfo{title}{{Light-induced evaporative cooling of holes in the
  {Hubbard} model}}.
\newblock \emph{\bibinfo{journal}{Nat. Commun.}} \textbf{\bibinfo{volume}{10}},
  \bibinfo{pages}{1--7} (\bibinfo{year}{2019}).
\newblock \urlprefix\url{https://doi.org/10.1038/s41467-019-13557-9}.

\bibitem{werner2019prb}
\bibinfo{author}{Werner, P.}, \bibinfo{author}{Li, J.},
  \bibinfo{author}{Gole\ifmmode~\check{z}\else \v{z}\fi{}, D.} \&
  \bibinfo{author}{Eckstein, M.}
\newblock \bibinfo{title}{Entropy-cooled nonequilibrium states of the {Hubbard}
  model}.
\newblock \emph{\bibinfo{journal}{Phys. Rev. B}}
  \textbf{\bibinfo{volume}{100}}, \bibinfo{pages}{155130}
  (\bibinfo{year}{2019}).
\newblock \urlprefix\url{https://link.aps.org/doi/10.1103/PhysRevB.100.155130}.

\bibitem{micnas1990}
\bibinfo{author}{Micnas, R.}, \bibinfo{author}{Ranninger, J.} \&
  \bibinfo{author}{Robaszkiewicz, S.}
\newblock \bibinfo{title}{Superconductivity in narrow-band systems with local
  nonretarded attractive interactions}.
\newblock \emph{\bibinfo{journal}{Rev. Mod. Phys.}}
  \textbf{\bibinfo{volume}{62}}, \bibinfo{pages}{113--171}
  (\bibinfo{year}{1990}).
\newblock \urlprefix\url{https://link.aps.org/doi/10.1103/RevModPhys.62.113}.

\bibitem{greif2011}
\bibinfo{author}{Greif, D.}, \bibinfo{author}{Tarruell, L.},
  \bibinfo{author}{Uehlinger, T.}, \bibinfo{author}{J\"ordens, R.} \&
  \bibinfo{author}{Esslinger, T.}
\newblock \bibinfo{title}{Probing nearest-neighbor correlations of ultracold
  fermions in an optical lattice}.
\newblock \emph{\bibinfo{journal}{Phys. Rev. Lett.}}
  \textbf{\bibinfo{volume}{106}}, \bibinfo{pages}{145302}
  (\bibinfo{year}{2011}).
\newblock
  \urlprefix\url{https://link.aps.org/doi/10.1103/PhysRevLett.106.145302}.

\bibitem{keiter1971}
\bibinfo{author}{Keiter, H.} \& \bibinfo{author}{Kimball, J.}
\newblock \bibinfo{title}{Diagrammatic perturbation technique for the
  {Anderson} {Hamiltonian}, and relation to the sd exchange {Hamiltonian}}.
\newblock \emph{\bibinfo{journal}{Int. J. Magnetism}}
  \textbf{\bibinfo{volume}{1}}, \bibinfo{pages}{233} (\bibinfo{year}{1971}).

\bibitem{eckstein2010prb}
\bibinfo{author}{Eckstein, M.} \& \bibinfo{author}{Werner, P.}
\newblock \bibinfo{title}{Nonequilibrium dynamical mean-field calculations
  based on the noncrossing approximation and its generalizations}.
\newblock \emph{\bibinfo{journal}{Phys. Rev. B}} \textbf{\bibinfo{volume}{82}},
  \bibinfo{pages}{115115} (\bibinfo{year}{2010}).
\newblock \urlprefix\url{https://link.aps.org/doi/10.1103/PhysRevB.82.115115}.

\bibitem{li2020mpl}
\bibinfo{author}{Li, J.}, \bibinfo{author}{Golez, D.}, \bibinfo{author}{Werner,
  P.} \& \bibinfo{author}{Eckstein, M.}
\newblock \bibinfo{title}{Superconducting optical response of photodoped {Mott}
  insulators}.
\newblock \emph{\bibinfo{journal}{Mod. Phys. Lett. B}}
  \textbf{\bibinfo{volume}{34}}, \bibinfo{pages}{2040054}
  (\bibinfo{year}{2020}).
\newblock \urlprefix\url{https://doi.org/10.1142/S0217984920400540}.

\end{thebibliography}
\section{\NoCaseChange{Acknowledgments}}
\begin{acknowledgments}
This project has received funding from the European Union’s Horizon 2020 research and innovation programme under the Marie Sk\l{}odowska-Curie grant agreement No. 884104, from ERC Consolidator Grant No.~724103 and from Swiss National Science Foundation Grant No.~200021-196966. JL thanks F.~Schlawin, T.~Kaneko, A.~Ramires, C.~Mudry, and M.~Eckstein for helpful discussions.
\end{acknowledgments}
\newpage
\appendix 
\section{\NoCaseChange{Methods}}
\textbf{Nonequilibrium dynamical mean-field theory solution.} We solve the driven Hubbard model on the Bethe lattice with infinite coordination number using nonequilibrium dynamical mean-field theory \cite{aoki2014}. The model can be exactly mapped to the Anderson model when the hopping is rescaled with $t_0/\sqrt{z}$ and $z\to\infty$ and solvable through dynamical mean-field theory, and thus provides a concrete solvable model to establish the $120^\circ$ order. This rescaling leads to a noninteracting bandwidth $4t_0$. The exchange interaction for each bond is given by $J_{\rm ex} =4t_0^2/zU$, while the total (mean-field) energy contribution scales with $z$ times this value, $4t_0^2/U$. This infinite-coordinational model captures local correlation effects and, in particular, the interplay between hopping and exchange interactions, while it completely neglects nonlocal correlations as well as the effects of electrons hopping around loops, present in lower dimensional systems. 

The lattice problem is exactly mapped to three single-impurity Anderson models on the Keldysh contour, defined by the action (spin index neglected for simplicity of notation)
\begin{align}
S_{X, \rm imp}&=\int dt c^\dag_X(t)(i\partial_t - h_{\rm loc})c_X(t)\nonumber\\
&-\int dt dt' c^\dag_X(t)[\Delta_X(t-t')+\sum_\ell D_{\ell}(t-t')]c_X(t'),
\end{align}
where $X=R,G,B$ and $\Delta_X(t,t')$ is the total bath hybridization function. For an $R$ site, half of its neighbors are $G$ sites and the other half are $B$, and similarly for the other two sites. The local Hamiltonian $h_{\rm loc}$ includes the Hubbard interaction and the pair seed term. The self-consistency relation, 
which is given below, 
yields $\Delta$ from the local Green's function $G_{\rm loc}(t,t')=G_{\rm imp}(t,t')$. $D_\pm$ is the bath hybridization. The steady-state problem is solved with a frequency-domain strong-coupling impurity solver, whose implementation is detailed in Ref.~\cite{li2021}. 

In the nonquilibrium steady-state setup,
the driving term reads $gH_{\rm dr}=\frac{g}{\sqrt{L}}\sum_{i\sigma\alpha} (c^\dag_{i\sigma} d_{i\alpha\sigma}+\text{h.c.})+\sum_{i\alpha\sigma}\epsilon_\alpha d^\dag_{i\alpha\sigma}d_{i\alpha\sigma}$ with bath operators $d_{i\alpha\sigma}$ and $\alpha=(\ell,\xi)$ containing a bath label $\ell=\pm$ and the energy level label $\xi$. $L$ represents the bath size and is of length dimension. Three impurity problems labelled by $R,G,B$ are solved with the non-crossing approximation (NCA) \cite{keiter1971,eckstein2010prb}.
In DMFT, the impurity hybridization function is determined by $\Delta_R(t,t')=t_0^2 \tau_z(e^{i\tau_z\varphi}G_B(t,t')e^{-i\tau_z\varphi}+e^{-i\tau_z\varphi}G_G(t,t')e^{i\tau_z\varphi})\tau_z/2$, and analogously for $\Delta_{G/B}$, maintaining a bandwidth of $4t_0$. 
The Pauli-matrix $\tau_z$ appears because of the Nambu formalism \cite{li2020}. In practice,  we apply a seed term $h\sum_i \theta_i c^\dag_{i\uparrow}c^\dag_{i\downarrow}+\text{h.c.}$ with $h=0.001$ to break the symmetry, where $\theta_i=1,e^{i2\pi/3},e^{i4\pi/3}$ for the $R,G,B$ site, respectively. The doublon number (per site) is calculated as $n_d=\langle n_\uparrow n_\downarrow\rangle$.

The real-time dynamics in Fig.~\ref{spec}(d,e) is obtained using the entropy-cooling protocol \cite{werner2019,werner2019prb}. To be concrete, the Hubbard system is coupled to two narrow bands at each site, which have a semielliptic DoS of half-bandwidth $0.1$ and are located at the energies $\omega_{\pm}=\pm 6$. The upper ($\omega_+$) band is empty while the lower band ($\omega_-$) is full, as the chemical potential is set to $\mu_b=0$. We drive the coupling constant $g=v(t)$ with a pulse as given below, inducing a resonant charge transfer between the empty (full) core level and the lower (upper) Hubbard band, respectively. The key idea is to adjust the frequency in time, so as to match the first Floquet sideband of the narrow band with the effective Fermi level for the doublons or holons (roughly speaking one wants $\omega_+-\Omega(t)\sim \mu_+(t)$, where $\mu_+$ is the Fermi level in the upper Hubbard band, and similarly for the lower Hubbard band). See Ref.~\citenum{werner2019prb} for more details. Specifically, the pulse is given by 
$v(t)=\frac{\sin[\Omega(t) t]}{(1+e^{(t-t_1-T)\gamma_{\rm off}})(1+e^{-(t-t_1)\gamma_{\rm on}})}$ for $t>0$ and zero otherwise, where $\Omega(t)=\Omega_i+(\Omega_f-\Omega_i)\sin(\pi t/400)$ and $t_1=4,T=100,\gamma_{\rm on}=1,\gamma_{\rm off}=1/4,\Omega_i=7.25,\Omega_f=12.5$. The frequency $\Omega(t)$ is varied to fill the upper Hubbard band to its top, and to empty the lower Hubbard band to its bottom. The parameters are empirically optimized to minimize $T_{\rm eff}$ in the final state and generate a long-lived order.

Within DMFT, the current flowing through site $R$ can be calculated by
\begin{align}
J&=-\frac{1}{2}\operatorname{Re}\tau_xG_R(t,t')*\Big[t_0^2 \tau_z(e^{i\tau_z\varphi}G_B(t,t')e^{-i\tau_z\varphi}\nonumber\\
&-e^{-i\tau_z\varphi}G_G(t,t')e^{i\tau_z\varphi})\tau_z/2\Big],
\end{align}
and similarly for the other sites. Here, the symbol $*$ represents a convolution on the Keldysh contour. 

\textbf{Exact diagonalization studies}. We solve the generalized $t$-$J$ model on the triangular and Kagome lattices with the artificial gauge field $\varphi$. Three-site terms are ignored, since they play a similar role as the electron hopping, assisting doublon/holon delocalization, but have a much lower strength $J_{\rm ex} \ll t_0$. Three conserved quantites, the number of up and down spins and the doublon number $n_d$, are imposed to satisfy $n_\uparrow=n_\downarrow=n_d$. For the triangular lattice, a 12-site cluster, as shown in Fig.~1(b), is solved with the Lanczos algorithm. We define $\bm a_1$ ($\bm a_2$) as in Fig.~\ref{diagram}(b), and the torus is spanned by $2\bm a_1+2\bm a_2$ and $2\bm a_1-4\bm a_2$. The green sites at the left and right bottom corners are not in the cluster, but are identified with sites in the cluster using the periodic boundary conditions. The pairing correlation is averaged over sites, using translational invariance and the structure factor is summed over chiralities (replacing $\theta_i\to \theta_i+\theta^*_i$). 

In the case of the Kagome lattice, a 12-site cluster, as shown in Fig.~1(c), is solved with the same method. The uppermost green sites and the rightmost red sites are identified with the lowermost and leftmost sites using periodic boundary conditions, respectively. We consider off-resonant polarized light in the long-wavelength limit. The hopping term along bond $\langle ij\rangle$ is then dressed with the Peierls phase $
\exp(i\bm r_{ij}\cdot \bm A(t))=\exp[i(\mathcal{A}_{ij} e^{i\Omega t}+\mathcal{A}^*_{ij}e^{-i\Omega t})],
$ where $\bm r_{ij} = \bm r_i - \bm r_j$ and $\mathcal{A}_{ij}=\mathcal{A}_0a e^{i\theta_{ij}}/2$ for the bond with lattice constant $a$ which is parallel to $(\cos\theta_{ij}, \sin\theta_{ij})$.

The $l$th Fourier component of the Hamiltonian reads
\begin{align}
H_l&=\delta_{l0}\sum_i U_in_{i\uparrow}n_{i\downarrow}-i^{|l|}t_0\sum_{\langle ij\rangle\sigma} e^{il\theta_{ij}}J_{|l|}(\mathcal{A})c^\dag_{i\sigma}c_{j\sigma},
\end{align}
where $\mathcal{A}=\mathcal{A}_0a/2$ the $J_l$ are Bessel functions of the first kind. In the high-frequency limit, one can obtain the effective Hamiltonian with a $1/\Omega$ expansion,
\begin{align}
 &H_0+\sum_{l>0} [H_l,H_{-l}]/l\Omega \nonumber\\
=& \sum_i U_in_{i\uparrow}n_{i\downarrow}-t_\text{R}\sum_{\langle ij\rangle\sigma} c^\dag_{i\sigma}c_{j\sigma}-t_{\rm NNN}\sum_{\llangle ijk\rrangle\sigma} c^\dag_{i\sigma}c_{k\sigma} \, .
\end{align}
The complex NN hopping and purely imaginary next NN hopping read 
\begin{align}
t_{\rm R}&=t_0 J_0(\mathcal{A})-it_0^2\sum_l(-)^lJ_l(\mathcal{A})^2\sin(2l\pi/3)/l\Omega,\nonumber\\
t_{\rm NNN}&=-it_0^2\sum_l(-)^lJ_l(\mathcal{A})^2\sin(l\pi/3)/l\Omega.
\label{floquet}
\end{align}
We include both hoppings in the simulation.

\onecolumngrid
\newpage
\begin{center}
\textbf{Supplementary Materials}
\end{center}
\section{$120^\circ$ twisted superconducting order at maximum photodoping}
\begin{figure*}[h]
\includegraphics[scale=1.3]{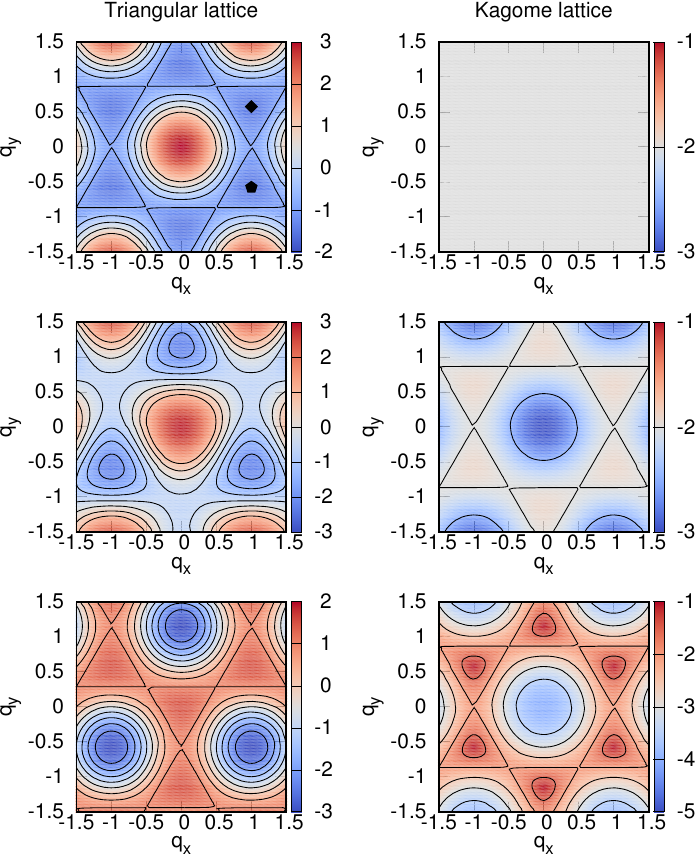}
\caption{The energy dispersion $\epsilon(\bm q)$ for orders with momentum $\bm q$ on the triangular and the Kagome lattices with $\varphi=0,\pi/20,\pi/6$ in the first, second, and third row, respectively. Note that the $q_{x,y}$ axes are the cartesian coordinates in the reciprocal space and have units of $2\pi/\sqrt{3}a$. The left panels show the energy dispersion for the triangular lattice. The top panel shows two minima at $\bm q=\pm[(2\pi/3)\bm b_1+ (-2\pi/3)\bm b_2]$, which are degenerate for $\varphi=0$ and marked by a black pentagon and diamond, respectively. One minimum, $\bm q=(2\pi/3)\bm b_1+ (-2\pi/3)\bm b_2$ (diamond), is selectively stabilized for $0<\varphi<\pi/3$. For $\varphi=\pi/6$ the stabilization is optimal, and the dispersion becomes nearly isotropic around the minimum, leading to the absence of leading-order trigonal warping and thus to the vanishing of the second-order supercurrent response. The right panels show the lowest-lying band of the Kagome lattice. For $\varphi=0$ the band is flat, while $\varphi>0$ stabilizes the $\bm q=0$ mode. }
\label{ek}
\end{figure*}

At maximum photodoping $n_d=1/2$, the effective Hamiltonian for the half-filled Hubbard model is given by the XXZ model
\begin{align}
H_{\rm eff}=\frac{J_{\perp}}{2}\sum_{\langle ij\rangle}(e^{2i\varphi_{ij}}\phi^+_i\phi^-_j+\text{h.c.})+J_{z}\sum_{\langle ij\rangle}\phi^z_i\phi^z_j,
\end{align}
where $\varphi_{RG}=-\varphi_{GR}=(\text{cyclic})=\varphi$ for uniform phases $\langle \phi^z_i\rangle=0$.  
We will investigate the condensate with ``momentum" $\bm q$, namely $\langle\phi^+_i\rangle=\phi_0e^{i\bm q\cdot \bm r_i}$.
On the triangular lattice, the energy per site is given by $\phi^2_0\epsilon(\bm q)$, where the dispersion $\epsilon(\bm q)$ reads
\begin{align}
\epsilon(\bm q)=J_{\perp}[\cos(q_1+2\varphi)+\cos(q_2-2\varphi)+\cos(q_2-q_1+2\varphi)].
\end{align}
The $q_{1,2}$ are the reciprocal coordinates, defined as $\bm q=q_1\bm b_1+q_2\bm b_2$, with reciprocal basis vectors $\bm b_1=(2/\sqrt{3},0),\bm b_2=(-1/\sqrt{3},1)$. The energy dispersion $\epsilon(\bm q)$ has two minima at $\bm q = \pm[(-2\pi/3)\bm b_1 + 2\pi/3\bm b_2)]$, for $\varphi=0$, corresponding to two different chiral $120^\circ$ condensates.  As $\varphi$ becomes nonzero, one of the two chiral states becomes the 
unique minimum in the energy landscape, with a reduced energy in the range $|\varphi|<\pi/3$. The situation for $\varphi>0$ is illustrated in the left panels of Fig.~\ref{ek}, where the minimum at $(2\pi/3)\bm b_1+(-2\pi/3)\bm b_2$ is stabilized.

In the case of the Kagome lattice, the energy dispersion splits into three bands because of the three atoms in a unit cell. A flat band emerges as the lowest-lying one at $\varphi=0$, and thus no particular order is favored. The actual ground state is selected by quantum fluctuations. However, if we impose the $\varphi_{ij}$ pattern shown in Fig.~1(c), the $\bm q=0$ order is stabilized as $\varphi$ increases. 

\subsubsection{Exactly solvable point}
For triangular and Kagome lattices, the Hamiltonian $H_{\rm eff}$ can be decomposed as $H_{\rm eff}=\sum_{\Delta} h(\Delta)$, i.e., into a sum of triangular motifs $\Delta$. In the following, we will prove that the $120^\circ$ condensate is the exact ground state for $J_{z}=J_{\perp}\cos(2\pi/3+\varphi)$ and $0<\varphi<\pi/3$, following the argument of Changlani {\it et al.} \cite{changlani2018}. For simplicity, we use pseudospin $\ket{\uparrow}$ and $\ket{\downarrow}$ to denote doublon and holon states, respectively. 

More percisely, we will prove that the coherent state of hard-core doublons 
\begin{align}
\ket{\Psi}=\exp(\sum_i\theta^*_i\phi_i^+)\ket{0}
\end{align}
is a ground state of the Hamiltonian if $J_{z}=J_{\perp}\cos(2\pi/3+\varphi)$. 
Here the sum is over all sites $i$, and the ``color" $\theta_i$ is defined as in the main text. This state is essentially a tensor-product state $\bigotimes_i \ket{C_i}$ with $C_i=R,G,B$ at each site $i$, defined as $\ket{R}=\frac{1}{\sqrt{2}}(\ket{\uparrow}+\ket{\downarrow}), \ket{G}=\frac{1}{\sqrt{2}}(\ket{\uparrow}+\omega\ket{\downarrow}),\ket{B}=\frac{1}{\sqrt{2}}(\ket{\uparrow}+\omega^2\ket{\downarrow})$ with $\omega=e^{i2\pi/3}$. 

The Hilbert space of each motif is spanned by eight orthogonal states, including the fully polarized $\ket{3/2}=\ket{\uparrow\uparrow\uparrow},\ket{-3/2}=\ket{\downarrow\downarrow\downarrow}$ and two sets of chiral states defined by
\begin{gather}
\begin{split}
\ket{+1/2}_L&=\ket{\downarrow\uparrow\uparrow}+\omega\ket{\uparrow\downarrow\uparrow}+\omega^2\ket{\uparrow\uparrow\downarrow},\\
\ket{-1/2}_L&=\ket{\uparrow\downarrow\downarrow}+\omega^2\ket{\downarrow\uparrow\downarrow}+\omega\ket{\downarrow\downarrow\uparrow},\\
\ket{+1/2}_R&=\ket{\downarrow\uparrow\uparrow}+\omega^2\ket{\uparrow\downarrow\uparrow}+\omega\ket{\uparrow\uparrow\downarrow},\\
\ket{-1/2}_R&=\ket{\uparrow\downarrow\downarrow}+\omega\ket{\downarrow\uparrow\downarrow}+\omega^2\ket{\downarrow\downarrow\uparrow},
\end{split}
\end{gather}
and two nonchiral states
\begin{gather}
\begin{split}
\ket{+1/2}_0&=\frac{1}{\sqrt{3}}(\ket{\downarrow\uparrow\uparrow}+\ket{\uparrow\downarrow\uparrow}+\ket{\uparrow\uparrow\downarrow})\\
\ket{-1/2}_0&=\frac{1}{\sqrt{3}}(\ket{\uparrow\downarrow\downarrow}+\ket{\downarrow\uparrow\downarrow}+\ket{\downarrow\downarrow\uparrow}).
\end{split}
\end{gather}
The above states are labelled by the total $z$-pseudospin $\phi^z$ (number of doublons) and the momentum $\pm 2\pi/3,0$ associated with the three-fold rotational symmetry. It is crucial to note that $\ket{\pm}_L$ are degenerate and $\ket{\pm}_R$ are degenerate due to the symmetry under a combined reflection and a particle-hole transformation (the exchange of $\ket{\uparrow}$ and $\ket{\downarrow}$). Indeed, the above six states are all eigenstates of the Hamiltonian with eigenvalues $\lambda_{\frac{3}{2},0}=\lambda_{-\frac{3}{2},0}=3J_{z}/4,\lambda_{\frac{1}{2},L}=\lambda_{-\frac{1}{2},L}=-J_{z}/4+J_{\perp}\cos(2\pi/3 +2\varphi),\lambda_{\frac{1}{2},R}=\lambda_{-\frac{1}{2},L}=-J_{z}/4+J_{\perp}\cos(2\pi/3 -2\varphi)$ and $\lambda_{\frac{1}{2},0}=\lambda_{-\frac{1}{2},0}=-J_{z}/4+J_{\perp}\cos(2\varphi)$, where we have labelled the eigenvalues by $\lambda_{\phi^z_{\rm tot}, c}$ with chirality $c=R,L,0$. We can verify this by explicit calculations. For $\ket{\pm}_{L}$ one has
\begin{gather}
\begin{split}
h(\Delta)\ket{+1/2}_L&=e^{-i2\varphi}\ket{\uparrow\downarrow\uparrow}+e^{i2\varphi}\ket{\uparrow\uparrow\downarrow}+e^{-i2\varphi}\omega\ket{\uparrow\uparrow\downarrow}+e^{i2\varphi}\omega\ket{\downarrow\uparrow\uparrow}+e^{-i2\varphi}\omega^2\ket{\downarrow\uparrow\uparrow}+e^{i2\varphi}\omega^2\ket{\uparrow\downarrow\uparrow}\\
&=\lambda_{\frac{1}{2},L}\ket{+1/2}_L\\
h(\Delta)\ket{-1/2}_L&=e^{i2\varphi}\ket{\downarrow\uparrow\downarrow}+e^{-i2\varphi}\ket{\downarrow\downarrow\uparrow}+e^{i2\varphi}\omega^2\ket{\downarrow\downarrow\uparrow}+e^{-i2\varphi}\omega^2\ket{\uparrow\downarrow\downarrow}+e^{i2\varphi}\omega\ket{\uparrow\downarrow\downarrow}+e^{-i2\varphi}\omega\ket{\downarrow\uparrow\downarrow}\\
&=\lambda_{-\frac{1}{2},L}\ket{-1/2}_L,
\end{split}
\end{gather}
which are similar for the chiral pairs. The single-motif Hamiltonian $h(\Delta)$ can then be recast into the following form,
\begin{align}
H=\sum_\Delta h(\Delta)&=\lambda_{\frac{1}{2},0}\sum_\Delta\mathcal{P}_{0}(\Delta)+\lambda_{\frac{1}{2},R}\sum_{\Delta}\mathcal{P}_R(\Delta)+\lambda_{\frac{1}{2},L}\sum_{\Delta}\mathcal{P}_L(\Delta)+\lambda_{\frac{3}{2}}\sum_\Delta\mathcal{P}_{3/2}(\Delta),
\end{align}
where $\mathcal{P}_{0,R,L,3/2}$ projects the states to the nonchiral, $R$, $L$, and fully polarized subspaces for a triangle $\Delta$, respectively. When $J_{z}=J_{\perp}\cos(2\pi/3+2\varphi)$, the fully polarized states and one set of chiral states $\ket{+1/2}_L$ and $\ket{-1/2}_L$ are degenerate as $\lambda_{\frac{1}{2},L}=\lambda_{\frac{3}{2}}$. For $0<\varphi<\pi/3$, the states of the opposite chirality and the nonchiral states lie higher in the energy spectrum of $h(\Delta)$. 

A key observation is that, for a single triangular motif, the three-coloring state reads
\begin{align}
\ket{R}\ket{G}\ket{B}&= \frac{1}{2^{3/2}}[\ket{+3/2}+\ket{-3/2}+(\ket{\uparrow\downarrow\downarrow}+\omega^2\ket{\downarrow\uparrow\downarrow}+\omega\ket{\downarrow\downarrow\uparrow})+(\ket{\downarrow\uparrow\uparrow}+\omega\ket{\uparrow\downarrow\uparrow}+\omega^2\ket{\uparrow\uparrow\downarrow})]\nonumber\\
&= \frac{1}{2^{3/2}}[\ket{3/2}+\ket{-3/2}+\ket{+1/2}_L+\ket{-1/2}_L],
\end{align}
which only contains fully polarized states and $\ket{\pm1/2}_L$.  As a result, at $J_{z}=J_{\perp}\cos(2\pi/3+2\varphi)$, the coherent state $\ket{\Psi}$ consistently zeros out $\mathcal{P}_R$ and $\mathcal{P}_0$, 
and must be a ground state with energy $3J_{z}N_\Delta/4$, where $N_{\Delta}$ is the number of triangular motifs.

This three-coloring state is an exact ground state for all $\phi^z$ sectors, thus equally favoring the $120^\circ$ condensate and charge segregation (uniform $\phi^z$). For the triangular lattice, when $|J_{z}|<|J_{\perp}\cos(2\pi/3+2\varphi)|$ one can argue that the $120^\circ$ order is favored. This parameter regime is probably most realistic in experiments due to the intersite Coulomb repulsion. We note that a relatively strong nearest-neighbor Coulomb repulsion $V\sum_{\langle ij\rangle}n_in_j$ can change $J_{z}$ to $J_{z}=V-4t_0^2/(U-V)>0$, favoring charge-density-wave order, which is frustrated on the lattices considered here. 

The stability against deviations from maximum photodoping ($n_d=1/2$) has been numerically studied in the main text. At $\varphi=\pi/6$, since the order commutes with the electron-hopping operator, one expects an extended range of stable $120^\circ$ condensate, at least when the energy scale of the hopping $n_s t_0$ is not much stronger than the exchange energy $J_{\rm ex}\sim 4t_0^2/U$, namely $n_s=1-2n_d\lesssim 4t_0/U$. 

\section{Superconducting optical response}
The doublon-holon condensate generally coexists with unpaired electrons, resulting in a gapless state with a large metallic conductivity and negative AC conductivity \cite{li2020mpl}. The superconducting current response, however, comes from the doublon-holon contribution, which can be expressed as $\mathcal{J}^{dh}_{ij}=\delta H_{dh}[\bm A]/\delta A_{ij}=i2J_{\perp}(e^{2i(\varphi_{ij}+A_{ij})}\phi^+_i\phi^-_j-\text{h.c.})$. 

For the triangular lattice, we now calculate the currents in the topmost triangular motif shown in Fig.~1(b). We assume that an electric pulse $\bm E(t)$ is applied up to time $t_f$ and $\varphi_{ij}=\varphi$ for a $B\to G\to R$ chain. When $t>t_f$, a constant vector potential $\bm A=-\int^{t_f}\bm E(s)ds$ has been generated and couples to the doublon-holon condensate. Assuming $\bm A=(A_x,A_y)$, we calculate $A_{ij}=\bm A\cdot (\bm r_i - \bm r_j)$ for the three bonds $\langle ij\rangle$ in the topmost triangular motif in Fig.~1(b),
\begin{gather}
\begin{split}
A_{RG}&=\frac{1}{2}A_y-\frac{\sqrt{3}}{2}A_x,\\
A_{GB}&=\frac{1}{2}A_y+\frac{\sqrt{3}}{2}A_x,\\
A_{BR}&=-A_y,
\end{split}
\end{gather}
where the lattice constant is set to unity. The three currents then read $\mathcal{J}_{ij}=\mathcal{J}^{dh}(A_{ij})$
for bond index $RG$, $GB$, and $BR$.

In general, we can express the current density as $\mathcal{J}^a=D^{ab}A_b+T^{abc}A_b A_c$ for $a,b,c=x/y$. The form of the coefficient tensors $D$ and $T$ is constrained by symmetry arguments. First of all, any rotational symmetry $C_n$ with $n\ge3$ requires $D^{ab}\propto \delta^{ab}$, since its faithful 2D representations are irreducible. Moreover, the $120^\circ$ condensate is invariant under reflection with respect to the $y$ axis, resulting in $T^{xyy}=T^{xxx}=T^{yxy}=0$. The invariance under the three-fold rotation further imposes $-T^{yyy}=T^{yxx}=T^{xxy}$.

We now concentrate on the bulk response of the triangular lattice. Note that the gauge-invariant current density generally reads $\bm J=\delta H/\delta\bm A$. On the mean-field level, the expectation value of the Hamiltonian per unit cell can be written as
\begin{align}
\frac{\langle H\rangle}{N_{\rm site}S_{\rm u.c.}}=\frac{\epsilon(\bm q_0+\bm A)}{S_{\rm u.c.}}=\frac{J_{\perp}\phi_0^2}{S_{\rm u.c.}}[\cos(2\pi/3+2\varphi+2A_{RG})+\cos(2\pi/3+2\varphi+2A_{GB})+\cos(2\pi/3+2\varphi+2A_{BR})],
\end{align}
where $\bm q_0=(2\pi/3,-2\pi/3)$ and the area of the unit cell is $S_{u.c.}=\sqrt{3}/2$. The energy stationarity guarantees that the net current density $\propto \partial \epsilon/\partial \bm A$ vanishes. The linear response is then determined by the concavity $D^{ab}\propto\partial^2 \epsilon/\partial A_a\partial A_b$ and the second-order response can be attributed to the trigonal warping $T^{abc}\propto\partial^3 \epsilon/\partial A_a\partial A_b\partial A_c$.

To obtain some physical insights, we calculate explicitly the current density,
\begin{gather}
\begin{split}
\mathcal{J}_x&=\frac{\partial \epsilon(\bm A)}{\partial A_x}\frac{1}{S_{\rm u.c.}}=\mathcal{J}_{GB}-\mathcal{J}_{RG},\\
\mathcal{J}_y&=\frac{\partial \epsilon(\bm A)}{\partial A_y}\frac{1}{S_{\rm u.c.}}=(\mathcal{J}_{RG}+\mathcal{J}_{GB}-2\mathcal{J}_{BR})/\sqrt{3}.
\end{split}
\end{gather}
These results can also be obtained by calculating the charge passing through a line perpendicular to the $x$- and $y$-directions per unit time and length. By expanding the above formula to second order, we obtain
\begin{gather}
\begin{split}
\mathcal{J}_x&\approx\frac{\partial \mathcal{J}^{dh}}{\partial A}\bigg|_{\bm A=0}\sqrt{3}A_x+\frac{\partial^2 \mathcal{J}^{dh}}{\partial A^2}\bigg|_{\bm A=0}\frac{\sqrt{3}}{2}A_xA_y,\\
\mathcal{J}_y&\approx\frac{\partial \mathcal{J}^{dh}}{\partial A}\bigg|_{\bm A=0}\sqrt{3} A_y+\frac{\partial^2 \mathcal{J}^{dh}}{\partial A^2}\bigg|_{\bm A=0}\frac{\sqrt{3}}{4}(A_x^2-A_y^2).
\end{split}
\end{gather}
When $A_y=0$, one can see that $\mathcal{J}_x\propto A_x$ while $\mathcal{J}_y$ does not vanish but is proportional to $A_x^2$. This gives rise to a second-order transverse current when $\bm A=(A_x,0)$. The $D$ and $T$ tensors are 
\begin{gather}
\begin{split}
D^{xx}&=D^{yy}=2\sqrt{3}J^{dh}_0\cos(2\pi/3+2\varphi),\\
T^{yxx}&=-T^{yyy}=T^{xxy}=-\frac{\sqrt{3}}{2}J^{dh}_0\sin(2\pi/3+2\varphi),
\end{split}
\end{gather}
from which one sees that the $T$ tensor vanishes for $\varphi=\pi/6$. The vanishing of the second-order response can be intuitively seen from Fig.~\ref{ek}, where the energy dispersion near the minimum becomes nearly isotropic which forbids a nonzero third-order derivative.

For the $\bm q=0$ order on the Kagome lattice, the pulse-induced supercurrent can become unbalanced and result in charge redistribution and thus an oscillation in the order parameter. For example, assuming $A_y=0,A_x\ne0$, the Blue site of the bottom left triangular motif in Fig.~\ref{kagome_orders}(a) has two incoming currents $\mathcal{J}^{dh}(-\sqrt{3}A_x/2)+\mathcal{J}^{dh}(\sqrt{3}A_x/2)$ and two outgoing $\mathcal{J}^{dh}(0)+\mathcal{J}^{dh}(0)$, differing by $\frac{\partial^2 \mathcal{J^{dh}}}{\partial A^2}3A^2_x/4$. Hence, an oscillation in the order parameter and the current is expected to be induced in this system. The $\sqrt{3}\times\sqrt{3}$ order, however, features a similar current response to the triangular lattice.

\section{Phase boundary and the critical doublon number for the Bethe lattice}
\begin{figure}[h]
\includegraphics[scale=0.6]{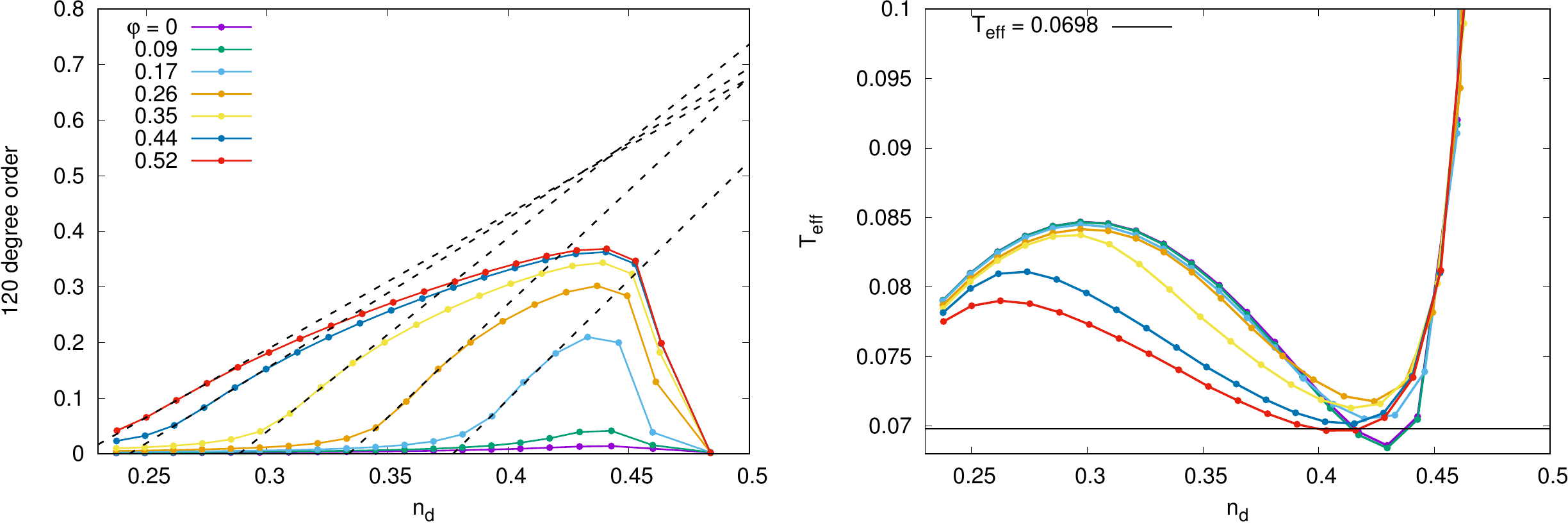}
\caption{Determination of the phase boundary by fitting the order parameter (left panel) and the effective temperature for $T_b=0.01$, $W=2$ (right panel). The effective temperature is estimated as the minimum of the curves with $\varphi>0.09$}
\label{bethe1}
\end{figure}

In this section, we show the fitting of the phase boundary and the effective temperature. The left panel of Fig.~\ref{bethe1} plots the order parameter as a function of $n_d$, which is varied by changing the bath chemical potential $\mu_b$. The data are fitted by dashed lines near the phase transition, and the critical doublon  densities $n^c_d$ are estimated as the vanishing points of the dashed lines. The transition is not sharply defined because of the weak symmetry-breaking field $h=0.001$ introduced in the DMFT iterations. For $W=2$, the effective temperature in the steady-state setup dramatically shoots up for $n_d\gtrsim0.43$, resulting in a strongly suppressed order parameter, and the phase boundary cannot be precisely determined in the above way. In this regime, both the external fermion baths and the upper/lower Hubbard bands of the system become close to full or empty, leading to a bottleneck for the thermalization and energy dissipation. This bottleneck can be partially overcome by using $W=2.7$. However, the intraband thermalization is generically reduced for a nearly full/empty band. 

We have also varied $T_b$ to obtain stationary states with different $T_{\rm eff}$, yielding a function $n^c_d(\varphi,T_{\rm eff})$. The $T_{\rm eff}$ generally varies along each curve for fixed $\varphi$, and as a conservative estimate, we have used the lowest value for $\varphi>0.09$ for all data points  
(see the horizontal line in the right panel of Fig.~\ref{bethe1}). This is a crude approximation which tends to underestimate $T_{\rm eff}$, and it thus guarantees that we get an \emph{lower bound} for the order paramter at a given $T_{\rm eff}$. The function $n^c_d(\varphi)$ is determined in the same way for each data set with different $T_b$. We have not fitted the $n^c_d$ for $\varphi=0$ and $\varphi=0.09$, since the order parameters are substantially smaller in these two cases and the fit becomes less robust. 

\begin{figure}[h]
\includegraphics[scale=0.6]{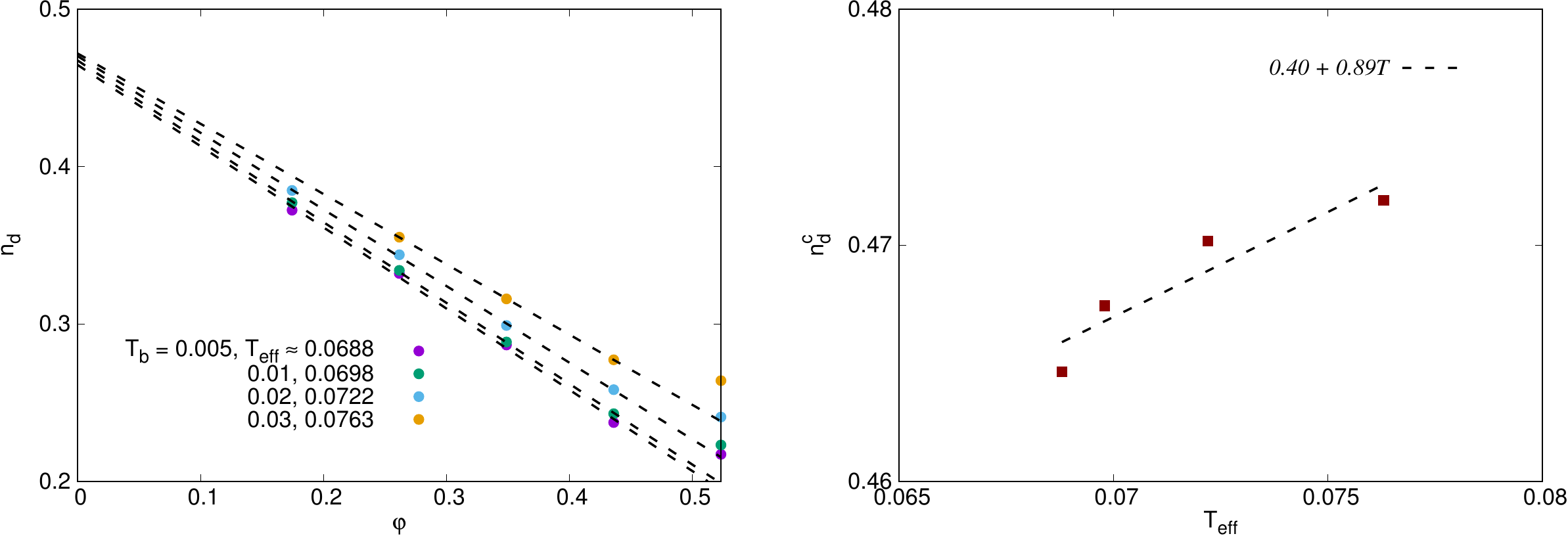}
\caption{Fitting of the phase boundary and temperature effects. The data are obtained by varying $T_b$ and the corresponding $T_{\rm eff}$ is estimated as in the text. The fermion bath has $W=2$. The left panel shows fits of the phase boundary which are extrapolated to $\varphi=0$. The extrapolated $\varphi=0$ critical doublon number is consistent with the order parameter obtained with the $W=2.7$ bath. The right panel shows the extrapolated $n_d(\varphi=0)$ as a function of $T_{\rm eff}$. From these values we can obtain the rather conservative upper bound $n_d\approx 0.40$ at $T_{\rm eff}\to0$.}
\label{fit}
\end{figure}

To estimate the critical doublon density at $T_{\rm eff}=0$ and $\varphi=0$, we first fit the curves $n^c_d(\varphi,T_{\rm eff})$ for each $T_b$ (with the estimated $T_{\rm eff}$) and extrapolate to obtain an estimate of $n^c_d(0, T_{\rm eff})$, see the left panel of Fig.~\ref{fit}. The $n^c_d$ data points for varying $\varphi$ and $T_b=0.01$ have been shown in Fig.~\ref{spec}, and we can see that the fitted $n^c_d(0)$ is consistent with the data for $\varphi=0$ and $W=2.7$. We then extrapolate the function $n^c_d(0,T_{\rm eff})$ to $T_{\rm eff}=0$, as shown in the right panel of Fig.~\ref{fit}. 

\section{The driven Hubbard model on the Kagome lattice}
\begin{figure}
\includegraphics[scale=0.3]{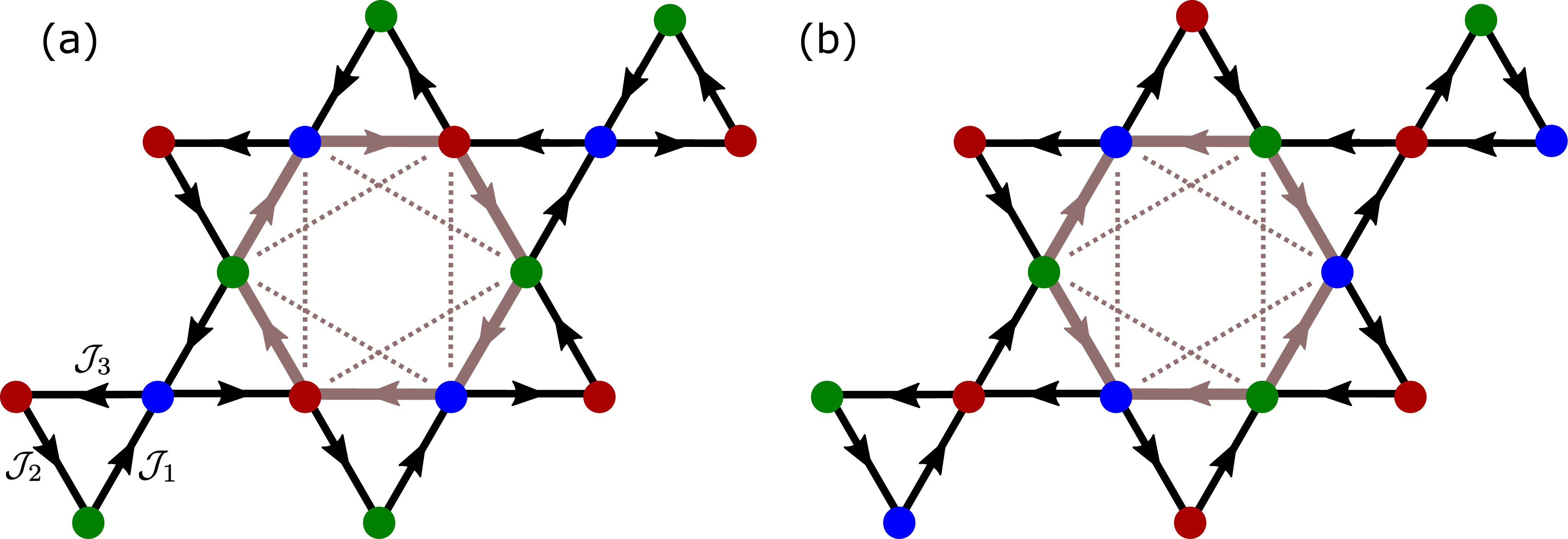}
\caption{Hubbard model on the Kagome lattice engineered by circularly polarized laser driving. The NN hopping is renormalized. The induced (second-order in $t_0$) next NN hopping is shown for one (light-brown) hexagon. 
 (a) When the NN hopping $t_\text{R}$ has a phase close to $\pi/6$, it favors the $\bm q=0$ order. (b) The next-nearest neighbor hopping favors identical order parameters on sites connected by the NNN bond, possibly favoring the uniform order or the $\sqrt{3}\times \sqrt{3}$ order shown here.}
\label{kagome_orders}
\end{figure}
\begin{figure}
\includegraphics[scale=0.7]{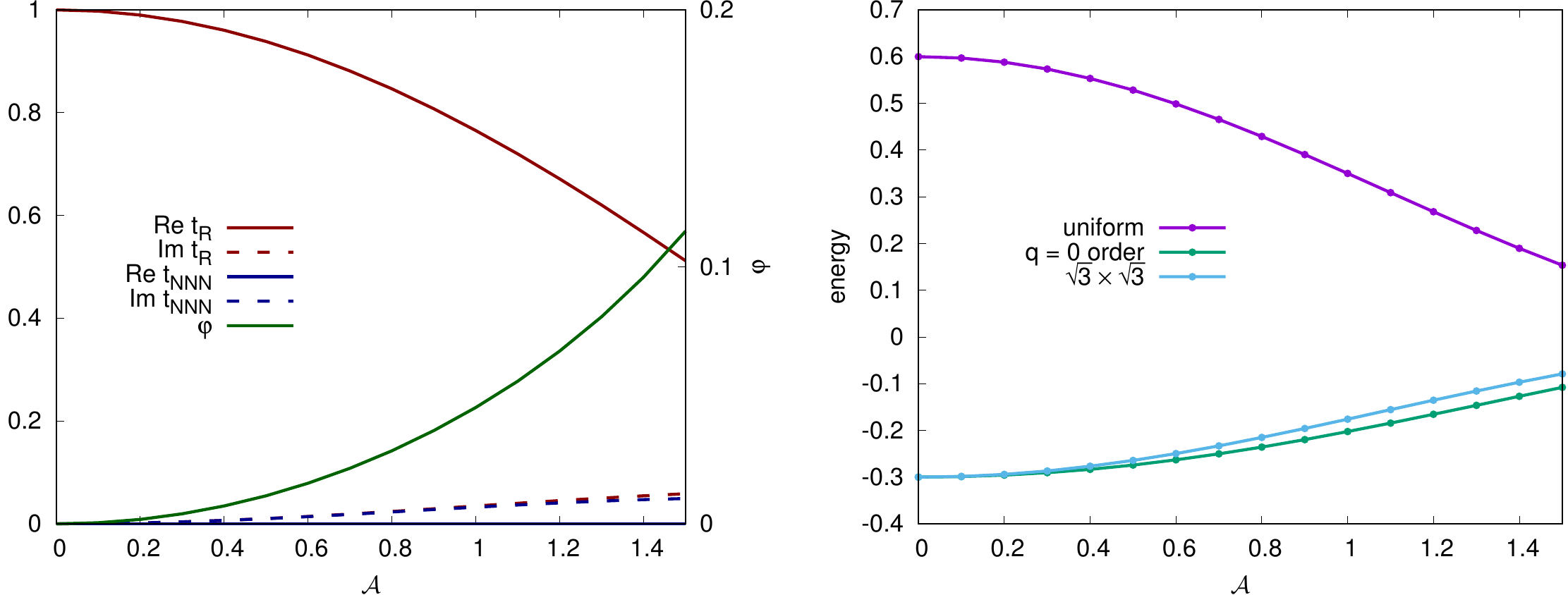}
\caption{The hopping parameters (left panel) and semiclassical energy (right panel) for different orders under driving, with $|\bm \phi|=0.5$ assumed. The parameters are $U/t_0=8, \Omega/t_0=5$. The $\bm q=0$ mode generally has the lower energy than the $\sqrt{3}\times\sqrt{3}$ state. In the strong driving limit, the phase of $t_\text{R}$ approaches $\pi/2$, and the uniform order becomes significantly enhanced.}
\label{kagome}
\end{figure}
Floquet driving with a circularly polarized light creates a complex phase for the hopping parameter $t_\text{R}$ up to the lowest order in a $1/\Omega $ expansion. In fact, it adds to the original $t_0$ a purely imaginary term at the order $t_0^2$ (see Methods). The resulting doublon-holon exchange term is of order $t_\text{R}^2/U\sim t^2_0/U + \mathcal{O}(t_0^3/U)$. Other terms generated by the light are of higher orders, including a purely imaginary next NN hopping at the order of $t_0^2$, and thus an exchange term of $\mathcal{O}(t_0^4/U)$. Therefore, we believe the main effect is from the renormalization of $t_\text{R}$. We also note that some Floquet-generated terms, such as the chiral spin interaction proportional to $\bm S_i\cdot(\bm S_j\times \bm S_k)$ \cite{claassen2017} can nevertheless alter the phase diagram of the system at the quantitative level.

In this section, we compare the mean-field energy of the uniform, the $\bm q=0$ and the $\sqrt{3}\times\sqrt{3}$ phases under various driving amplitudes $\mathcal{A}$, see Fig.~\ref{kagome_orders}. Specifically, we assume $\phi_0=1/2$ and evaluate the doublon-holon interaction term with a mean-field decoupling $\langle \phi^+_i\phi^-_i\rangle\to \phi_0^2\theta_i\theta^*_j$, where $\theta_i$ is defined as in the pairing structure factor. It suffices to calculate and compare the mean-field energy for a single hexagon. The driving-induced NN and NNN hoppings are shown in Fig.~\ref{kagome}(a) and the corresponding mean-field energies are shown in panel (b). The $\bm q=0$ $120^\circ$ order has the lowest energy among the three orders, consistent with the ED results shown in the main text.

\end{document}